\documentclass[longauth]{aa}  
\usepackage{graphicx}
\usepackage[varg]{txfonts}
\usepackage{booktabs}
\usepackage{xcolor}
\usepackage{xspace}
\usepackage{natbib,twoopt}
\usepackage{multirow}
\usepackage{longtable}
\usepackage{ltablex}
\usepackage{epsfig}
\usepackage{threeparttable}
\usepackage[breaklinks=true, colorlinks=true, allcolors=blue]{hyperref}
\newcommand{\nustar}{{\it NuSTAR}\xspace}
\newcommand{\nicer}{{\it NICER}\xspace}
\newcommand{\swift}{{\it Swift}\xspace}

\newcommand{\ep}{{\it Einstein Probe}\xspace}
\newcommand{\lco}{{\it LCO}\xspace}
\newcommand{\maxi}{{\it MAXI}\xspace}
\newcommand{\ergs}{erg s$^{-1}$\xspace }

\newcommand{\src}{Aql X-1\xspace}
\newcommand\T{\rule{0pt}{2.2ex}}       
\newcommand\B{\rule[-0.8ex]{0pt}{0pt}}

\def\flux{\mbox{erg\,cm$^{-2}$\,s$^{-1}$}}

\begin{document} 
\title{Aql X-1 \emph{from dawn 'til dusk:} the early rise, fast state transition and decay of its 2024 outburst }
\author{A. Marino\inst{1,2,3}\thanks{\url{marino@ice.csic.es}}, F. Coti Zelati\inst{1,2,4}, K. Alabarta\inst{4}, D. M. Russell\inst{5}, Y. Cavecchi\inst{6,7,8,9}, N. Rea\inst{1,2}, S. K. Rout\inst{5}, T. Di Salvo\inst{10}, J. Homan\inst{11}, Á. Jurado-López\inst{1}, L. Ji\inst{12}, R. Soria\inst{13,14,15}, T. D. Russell\inst{3}, Y. L. Wang\inst{15,1,2,16}, A. Anitra\inst{17,10}, M. C. Baglio\inst{4}, H. Feng\inst{18}, S. Fijma\inst{19,20}, S. Guillot\inst{21}, Y. F. Huang\inst{22,23}, G. Illiano\inst{4,1,2}, M. Imbrogno\inst{1,2,24}, C. Jin\inst{16}, F. Lewis\inst{25,26}, Y. F. Liang\inst{27,28}, M. J. Liu\inst{16}, R. Ma\inst{29}, G. Mastroserio\inst{30}, S. E. Motta\inst{4}, J. U. Ness\inst{31}, E. Parent\inst{1,2}, A. Patruno\inst{1,2}, P. Saikia\inst{32}, L. Tao\inst{19}, M. Veresvarska\inst{1,2}, X. P. Xu\inst{16,15}, W. Yuan\inst{16,15}, G. B. Zhang\inst{33}, Z. J. Zhang\inst{34,35}}

\institute{Institute of Space Sciences (ICE, CSIC), Campus UAB, Carrer de Can Magrans s/n, E-08193 Barcelona, Spain
         \and
             Institut d'Estudis Espacials de Catalunya (IEEC), 08860 Castelldefels (Barcelona), Spain 
             \and
            INAF/IASF Palermo, via Ugo La Malfa 153, I-90146 - Palermo, Italy 
              \and 
            INAF–Osservatorio Astronomico di Brera, Via Bianchi 46, I-23807 Merate (LC), Italy
            \and
            Center for Astrophysics and Space Science (CASS), New York University Abu Dhabi, PO Box 129188, Abu Dhabi, UAE
            \and
            Departamento de Astrof\'isica, Universidad de La Laguna, 38206, San Crist\'obal de La Laguna, Tenerife, Spain
            \and
            Instituto de Astrof\'isica de Canarias, 38205, San Crist\'obal de La Laguna, Tenerife, Spain
            \and
            Departament de Fis\'{i}ca, EEBE, Universitat Polit\`ecnica de Catalunya, Av. Eduard Maristany 16, 08019 Barcelona, Spain
            \and
            Center for Nuclear Astrophysics across Messengers (CeNAM), 640 S Shaw Lane, East Lansing, MI 48824, USA
            \and
            Dipartimento di Fisica e Chimica - Emilio Segr\`e, Universit\`a di Palermo, via Archirafi 36 - 90123 Palermo, Italy
            \and
            Eureka Scientific, Inc., 2452 Delmer Street, Oakland, CA 94602, USA 
            \and 
            School of Physics and Astronomy, Sun Yat-sen University, Zhuhai, 519082, People's Republic of China
            \and
            INAF – Osservatorio Astrofisico di Torino, Strada Osservatorio 20, I-10025 Pino Torinese, Italy
            \and 
            Sydney Institute for Astronomy, School of Physics A28, The University of Sydney, Sydney, NSW 2006, Australia
            \and
            School of Astronomy and Space Science, University of Chinese Academy of Sciences, 19A Yuquan Road, Beijing 100049, China
            \and
            National Astronomical Observatories, Chinese Academy of Sciences, 20A Datun Road, Beijing 100101, China
            \and
             Dipartimento di Fisica, Universit\`a degli Studi di Cagliari, SP Monserrato-Sestu, KM 0.7, Monserrato, 09042 Italy
            \and
            Key Laboratory of Particle Astrophysics, Institute of High Energy Physics, Chinese Academy of Sciences, Beijing 100049, People's Republic of China
            \and
            Anton Pannekoek Institute for Astronomy, University of Amsterdam, Science Park 904, 1098 XH, Amsterdam, the Netherlands
            \and
            ESO, Karl-Schwarzschild-Strasse 2, D-85748 Garching bei München, Germany
            \and
            IRAP, CNRS, 9 avenue du Colonel Roche, BP 44346, 31028 Toulouse Cedex 4, France;
            \and
            School of Astronomy and Space Science, Nanjing University, Nanjing 210023, China
            \and
            Key Laboratory of Modern Astronomy and Astrophysics (Nanjing University), Ministry of Education, Nanjing 210023, China
            \and
            INAF--Osservatorio Astronomico di Roma, via Frascati 33, I-00078 Monteporzio Catone, Italy
            \and
            Faulkes Telescope Project, School of Physics and Astronomy, Cardiff University, The Parade, Cardiff, CF24 3AA, Wales, UK
             \and
             The Schools’ Observatory, Astrophysics Research Institute, Liverpool John Moores University, 146 Brownlow Hill, Liverpool UK
            \and
            Purple Mountain Observatory, Chinese Academy of Sciences, Nanjing 210023, China
            \and
            School of Astronomy and Space Sciences, University of Science and Technology of China, Hefei 230026, China
            \and
            School of Physics and Astronomy, University of Southampton, Southampton, SO17 1BJ, UK
            \and
            Scuola Universitaria Superiore IUSS Pavia, Piazza della Vittoria 15, I-27100, Pavia, Italy
            \and 
            European Space Agency (ESA), European Space Astronomy Centre (ESAC), Camino Bajo del Castillo s/n, E-28692 Villanueva de la Cañada, Madrid, Spain
            \and
            Department of Astronomy, Yale University, PO Box 208101, New Haven, CT 06520-8101, USA
            \and
            Yunnan Observatories, Chinese Academy of Sciences, Kunming 650216, People's Republic of China
            \and
            Department of Physics, The University of Hong Kong, Pokfulam Road, Hong Kong SAR, China
            \and
            The Hong Kong Institute for Astronomy and Astrophysics, The University of Hong Kong, Hong Kong SAR, China
      }
  
\date{XXX-XXX}
\abstract{Transient Low-Mass X-ray Binaries (LMXBs) are usually first detected by all-sky X-ray monitors when they enter new outbursts. These detections typically occur at X-ray luminosities above $\sim$10$^{36}$ erg/s. As such, observations of these sources during the early rise of the outbursts have so far been very limited. However, the launch of the \emph{Einstein Probe} (EP) has greatly improved our ability to detect fainter X-ray activity, unlocking access to the outburst early rise. In September 2024, EP detected the early onset of a new outburst from the well-known neutron star LMXB Aql X-1, catching the source at a luminosity below 10$^{35}$ erg/s. In this paper we present results from a comprehensive, multi-wavelength campaign of this event, combining data from EP, \nicer, \nustar, \swift\ and Las Cumbres Observatory covering the full outburst from its early rise through its return to quiescence. By comparing X-ray and optical light curves obtained with Las Cumbres Observatory during the initial rise, we show that the start of the X-ray emission lagged the optical rise by, at most, 13 days, although,  as we did not observe the exact moment where the source goes
from (optical or X-rays) quiescence to rise, other lags cannot be completely ruled out. Time-resolved X-ray spectroscopy revealed how the geometry and the physical properties of the accretion flow evolve during this early stage of the outburst, as well as at higher luminosities as the source transitioned through the canonical X-ray spectral states - hard, intermediate and soft. These data show that the source underwent a very rapid, about 12-h long, transition from the hard state to the soft state about two weeks after the optical onset of the outburst. At the state transition, peculiar trends are observed for the temperature and physical sizes of both the inner region of the disk and a blackbody near the NS surface, which could suggest that at this stage a geometrically thicker inner disk emerges. We discuss these results in the context of time-scales for outburst evolution and state transitions in accreting neutron stars and black holes.}
\keywords{ Stars: neutron -- accretion, accretion disks -- X-rays: binaries -- X-rays: individual: \src}
\authorrunning{A. Marino}
\titlerunning{Aql X-1 from early rise to decay}
\maketitle

\section{Introduction}
Low-mass X-ray binaries (LMXBs) are binary systems composed of a black hole (BH) or a neutron star (NS) in orbit with a companion star of mass typically $\lesssim1$~M$_\odot$ \citep{Bahramian2023}. The electromagnetic output from these objects is powered by accretion of matter transferred from the companion star onto the compact object in the form of an accretion disk. The disk emission comprises a wide range of wavelengths, from near-IR and optical (originating in the outer disk) to hard X-rays, at hundreds of keVs (coming from the innermost region of the accretion flow).  During their weeks-to-months long transient outbursts, LMXBs in outburst brighten by several orders of magnitude at all wavelengths. In particular, the X-ray luminosity can reach values close to the Eddington limit, attaining levels around 10$^{38}$-10$^{39}$ erg/s. Outside of these episodes, LMXBs are thought to reside in a regime called quiescence at X-ray luminosities below 10$^{33}$ erg/s, where accretion onto the compact object proceeds at a much slower rate. 

Outbursts have been studied in detail from both an X-ray spectral and timing point of view, providing insights into how the geometry and the physical properties of the accretion flow change over time. Typically, three main components can be identified in the X-ray spectra of LMXBs \citep{Lin2007, DiSalvo2023}: a multi-color disk blackbody originating from the inner regions of the disk; a Comptonization component characterized by power-law index $\Gamma$ of $\sim$1.5-3.0 and energy cut-off in the range 1-100 keV or sometimes above and originating from an optically thin hot corona; and  a reflection spectrum \citep[e.g.][]{Fabian1989} due to the reprocessing of the emission from the corona by the disk. In accreting NSs, an additional blackbody component can be emitted from either the NS surface or a boundary layer (BL) in between the inner disk and the compact object \citep[e.g.][]{Inogamov1999,Babkovskaia2008, Suleimanov2006, Marino2023}. Depending on the broadband spectral shape, three main spectral states can be identified in LMXBs: a hard state, with emission dominated by the Comptonization component, a soft state, dominated by the disk (and/or the NS/BL components in NS LMXBs) and an intermediate state in between the two. The origin of these states has been historically associated to different stages of truncation of the optically thick disk \citep[e.g.][]{Esin1997}, with the hard state being the regime where the disk is truncated furthest from the compact object \citep[see, e.g.][and references therein for the debate regarding this model]{Zdziarski2020}.  As they evolve through outburst, LMXBs often exhibit all three states in a hysteresis cycle from hard to intermediate to soft state and then backwards \citep[e.g.][]{Homan2001,Fender2004,Dunn2010,Marcel2018a}. When plotted in a Hardness Intensity Diagram (HID), these sources typically trace a q-shaped track, so that these spectral states cycles are typically dubbed "q-tracks" or "q-diagrams" \citep[e.g.][]{Miyamoto1995,Zdziarski2004,Fender2004,Fender2009}. While q-tracks are observed in both BH and NS LMXBs, the two classes of sources occasionally show markedly distinct behaviors; while BH sources generally perform horizontal hard-to-soft transition over time-scales of weeks \citep[remaining at an approximately constant X-ray luminosity, e.g.][]{Tetarenko2016review}, in NS LMXBs these transitions are diagonal, entailing a rise in X-ray luminosity, and occur rapidly, over time-scales of days \citep[e.g.][]{MunozDarias2014,Marino2019b}.  Similarly to the energy spectra, the observed X-ray Fourier power density spectra show a dramatic evolution throughout the q-track, with strong broadband noise only in the hard state and a significant drop of X-ray variability as the source moves to softer states \citep[see, e.g.,][]{Vanderklis2006_review}. 

However, the picture painted above describes only the evolution of LMXBs at luminosities above 1\% of the Eddington limit, i.e., when they reach an X-ray luminosity of $\gtrsim10^{36}$ erg/s. The early rise of an outburst, where systems brighten from quiescence, can be considered as an uncharted regime due to the limitations of current all-sky monitors. Past and present all-sky X-ray monitors, such as those on board the {\it Rossi X-ray Timing Explorer}, the {\it Monitor of All-sky X-ray Image} (\maxi/GSC), and the {\it Neil Gehrels Swift Observatory}, have not been sensitive enough to catch new X-ray activity unless it reached a triggering luminosity of about 3$\times$10$^{36}$ erg/s for Galactic sources or higher, i.e., when the source had already entered the outburst phase completely. The early study of LMXB outbursts has so far significantly relied on continuous monitoring through ground-based optical telescopes, in particular through the program X-ray Binary New Early Warning System \citep[XB-NEWS][]{Russell2019} and similar efforts \citep[e.g.][]{Hameury1997,Jain2001,Buxton2004}. Due to these observational limitations, the physical mechanism responsible for triggering an outburst remains largely unclear. 

The canonical model used to explain transient LMXBs has historically been the Disk Instability Model \citep[DIM;][]{Osaki1974,Lasota2001}, which was initially invoked to explain outbursts in cataclysmic variables and then extended to accreting NSs and BHs \citep[see][for a recent review of the DIM in LMXBs]{Hameury2019}. According to such theory, outbursts are initiated from thermal-viscous instabilities starting as the cold disk in quiescence accumulates mass from the companion star. The increase in temperature and ionization caused by the instability will then propagate throughout the disk as a heating front, to the point where the cold quiescent disk transitions to a hot ``outbursting'' one and accretion onto the compact object is established. The heating front will, however, activate different parts of the disk at different times, causing a delay between the optical (originating from the outer disk) and the X-ray (from the inner disk and the Comptonizing hot flow) light curves of an outburst. However, optical-to-X-ray delays have been measured only for a handful of sources, mostly BH LMXBs \citep[e.g.][]{Orosz1997, Jain2001, Zurita2006,Bernardini2016,Tucker2018} and a few NSs, such as Aquila X$-$1 \citep[][]{Russell2019}, SAX J1808.4$-$3658 \citep{Goodwin2020} and  MAXI J1807+132 \citep{Rout2025_1807}. In all these cases, the optical start of the outburst has been observed to precede the X-ray start by several days.

The chances to catch the onset of new LMXB outbursts in the X-rays have significantly increased with the launch of the \ep \citep[EP, ][]{Yuan2025} mission. Thanks to the unprecedented ``grasp'' (field of view times sensitivity) of the Wide-field X-ray Telescope \citep[WXT, ][]{Cheng2025_wxt}, at least one order of magnitude larger than current and past X-ray wide-field monitors, and in the soft X-ray range, EP is able to detect X-ray Galactic sources with a 0.5--10 keV luminosities of about 10$^{35}$ erg/s (assuming an 8~kpc distance) or lower. In its 1.5 years of operations, EP has indeed been the first telescope to catch the onset of new outbursts from several known LMXBs \citep{Xu2024_atel, Wang2025_atel, Sun2025_atel}, including the transient NS LMXB Aquila X-1 \citep{Liu2024_atel}, hereafter \src.  

Aql X-1 (V1333 Aql) is one of the first NS LMXBs ever discovered, with its identification dating back to 1973 \citep{Kunte1973}. Thanks to its proximity, at a distance of  6$\pm$2 kpc \citep{Galloway2008,Matasanchez2017}, and its frequent, almost yearly, outbursts \citep{Simon2002, Niwano2023,Heinke2025}, Aql X-1 has been studied in great detail over the last few decades, becoming one of the archetypal transient NS LMXBs. The donor star has been identified as a K-type star \citep{Matasanchez2017} orbiting around the NS in a 18.95 hr orbit \citep{Chevalier1991}. Type-I X-ray bursts have been frequently observed from the source \citep[see, e.g.,][]{Mandal2025}, providing a secure identification of the accreting compact object as a NS \citep{Galloway2008} and of the NS spin frequency through the detection of burst oscillations \citep[549 Hz,][]{Zhang1998_boscill}. The source is also classified as an intermittent accreting millisecond pulsar \citep{DiSalvo2022}, briefly displaying X-ray pulsations at comparable frequency but only over a brief 150-s time interval \citep{Casella2008}. During its outbursts, Aql X-1 displays a canonical hysteresis cycle, undergoing transitions through the main known LMXBs spectral states \citep{Maitra2004, Guver2022, Putha2024}. In the last two years, the source underwent two outbursts, a bright one in 2024, discussed in this paper, and one significantly fainter in 2025 \citep{Alabarta2025_atel_aqlx1,Rout2025_atel_aqlx1}. The 2024 bright outburst of the source was discovered by WXT onboard EP on September 14th, 2024 \citep{Liu2024_atel}. The onset of the outburst was confirmed at optical wavelengths by XB-NEWS later that same day \citep{Rout2024_atel}. Multi-band follow-ups were performed in the following weeks with the Very Large Array \citep{Russell2024_atel} and MeerKAT \citep{Grollimund2024_atel} in radio and with \swift \citep{Mandal2024_atel} and SVOM/ECLAiRs \citep{LeStum2024_atel} in X-rays. 

In this work, we analyse the 2024 outburst of \src with \ep, \nicer, \nustar, \swift and {\it Las Cumbres Observatory} data. In Section \ref{sec:obs}, we report on the data reduction procedure used for all the aforementioned observatories. A detailed description of the data analysis, including a multi-band analysis of the outburst light curves (Section \ref{ss:lcurves}), a broadband X-ray spectral analysis (Section \ref{ss:spectral}) and an X-ray timing analysis (Section \ref{ss:timing}) are illustrated in Section \ref{sec:data}. Finally, Section \ref{sec:disc} is devoted to a discussion of the results presented in the previous sections, while conclusions and future prospects are presented in Section \ref{sec:concl}.

\section{Observations and Data Reduction}\label{sec:obs}
The outburst studied in this work started in early September 2024, lasting until mid-November 2024. Throughout this period, \src was the subject of a very dense monitoring campaign that provided almost daily coverage of the source for most of its active phase. In this paper we use data from several X-ray instruments as well as optical data, as mentioned above. In the following Sections, we illustrate the data reduction procedure employed for each instrument. As a thorough analysis of the Type-I X-ray bursts activity goes beyond the scope of this paper, we have systematically excluded all the bursts from the Good Time Intervals used for each dataset considered here. Details on all the X-ray observations used in this work are reported in Table \ref{tab:spec-log}. All X-ray data reduction processes have been carried out with \texttt{HeaSOFT} v. 6.33.1. 

\subsection{\ep}\label{ss:ep}
The 2024 outburst was first discovered by WXT onboard EP on September 14th, 2024 \citep{Liu2024_atel}. The source location has been covered daily by the monitor over a six-month period starting in July 2024 and ending on October 10th 2024, when unfortunately the source became not visible, thus interrupting the WXT monitoring of the 2024 outburst. We have gathered all WXT data from September 8th, 6-days prior to the detected outburst onset, until the last day that the source was visible. The total exposure of the accumulated WXT data is about 0.4 Ms. In order to extract a WXT light curve, we ran the processing pipeline available within the WXT Data Analysis Software (\texttt{WXTDAS}) to produce cleaned event files for each observation. A circular, 9\arcmin\-radius region centred on the source position was used to extract source photons. A region of the same size, but located away from the source, was used as the background region. We finally used \texttt{Xselect} to extract background-subtracted light curves from the cleaned event files in the 0.5--4 keV energy band. 

After the discovery of the outburst, we triggered an X-ray campaign with the Follow-up X-ray Telescope (FXT) onboard EP, collecting 5 observations in total with an average 2-day cadence. The observations were performed with one detector (A) in Partial Window mode (PW), and the other (B) in Timing Mode (TM). However, the source quickly became too bright for the PW observations, causing severe pile up at fluxes higher than 40 mCrab. By visually inspecting the pattern distribution plots created with the \texttt{fxtplotgrade} task, we checked whether the pile-up effect could be effectively mitigated using annular regions with increasingly larger inner radii, without success. As such, in the following we only consider the data taken in TM, which is significantly less impacted by pile-up. Rectangular regions of $3 '$ length and $1 '$ width aligned with the window are used for both source and background extractions. Further processing of the data is performed with the FXT Data Analysis Software Package (FXTDAS v.1.20) and \texttt{Xselect}. As the count-rate was visibly increasing within obsID 06800000125 (MJD 60581.692), we broke down the observation into two segments and extracted science products separately. 

\subsection{\nicer}\label{ss:nicer}
\nicer extensively monitored the 2024 outburst, with near-daily coverage totaling 44 observations. These data were processed using the \texttt{nicerl2} pipeline and default screening settings. In particular: i) we excluded time intervals in the proximity of the South Atlantic Anomaly; we considered only data taken with ii) elevation angle of at least 30$^\circ$ over the Earth's limb; iii) minimum angle of 40$^\circ$ from the bright Earth limb; iv) maximum angular distance between the source direction and \nicer pointing direction of 0.015$^\circ$. We note that the use of \texttt{HEASoft} v6.33.1 substantially mitigated the impact of the optical leak reported in May 2023. We inspected each light curve to check for the presence of Type-I X-ray bursts or other flares of non-physical nature, such as overshoots caused by charged particle contamination\footnote{\url{https://heasarc.gsfc.nasa.gov/docs/nicer/analysis_threads/overshoot-intro/}}. The intervals corresponding to these episodes were excluded from further analysis. Finally, some observations showed significant variability from one orbit to another. In those cases, we extracted science products separately for each orbit. In particular, this was the case for obsIDs: 7050340108, 7050340109, 7050340110 and 7675010106, which were broken down in 3, 3, 3 and 2 segments, respectively. We produced the spectra and background files with the \texttt{nicerl3-spect} pipeline, using the \textsc{scorpeon} default model for the creation of the background file. 

\subsection{\nustar}\label{ss:nustar}
\nustar observed \src twice during the 2024 outburst, on September 23rd and October 19th, for a total exposure of 38.5 ks. We analysed the data using standard tools provided by the \texttt{NuSTARDAS} package (version \texttt{v.2.1.4a}). The source spectrum was extracted from a circular region with a 100\arcsec\ radius centered on the source coordinates. To mitigate potential background non-uniformity across the detector, background spectra were obtained from four circular regions, each approximately 50\arcsec\ in radius, positioned at various source-free locations within the field of view. Light curves were extracted using \texttt{Xselect} and visually inspected for the presence of Type-I X-ray bursts. To isolate the persistent emission, we used \texttt{NuPRODUCTS} to extract spectra and light curves from an event file in which any such bursts had been filtered out.

\subsection{\swift}\label{ss:swift}
The Ultra-Violet/Optical Telescope (UVOT; \citealt{Roming2005}) on board \swift\ observed \src\ multiple times during the early rise and peak of its 2024 outburst. UVOT provided coverage in all six filters ($v$, $b$, $u$, $uvw1$, $uvm2$, and $uvw2$), although not every filter was used in each observation. We reprocessed all data starting from the raw sky images and combined exposures from individual visits using the \texttt{uvotimsum} tool. Source magnitudes were extracted with \texttt{uvotsource}, adopting a circular extraction region with a 5\arcsec\ radius centered on the source. The background level was estimated using a nearby 10\arcsec\ radius circular region free of contaminating stars. Aperture corrections were applied using the default curve-of-growth method. For each observation, source magnitudes were measured in the Vega system, and statistical and systematic uncertainties were combined in quadrature.  When the source was not significantly detected in a given filter, we report the corresponding 3$\sigma$ upper limits. A complete summary of the measurements is provided in Table\,\ref{tab:multiband}.

\subsection{\it Las Cumbres Observatory}\label{ss:lco}
Aql X-1 is one of the sources regularly monitored with the Las Cumbres Observatory (LCO) 1-m and 2-m telescopes as part of the Faulkes Telescope Project\footnote{\url{http://www.faulkes-telescope.com}} \citep{Lewis2008}. Observations were performed using the $V$, $R$ Bessell filters and the SDSS $g^{\prime}$, $r^{\prime}$, $i^{\prime}$ and $z^{\prime}$ filters. The optical magnitudes were extracted after performing multi-aperture photometry (MAP; see \citealt{Stetson90}) using the ``X-ray Binary New Early Warning System'' (XB-NEWS) pipeline (\citealt{Russell2019}; \citealt{Goodwin2020}; Alabarta et al, in prep).  XB-NEWS first detects sources in each image using SExtractor \citep{Bertin1996} and then gets the astrometric solution with astrometry.net \citep{Lang2010}, matching the detected sources to Gaia DR2 positions. If the target is not detected within 1\arcsec\ of its known coordinates at the default threshold, the pipeline re-runs using a lower threshold or forced photometry is performed at the target position. Then, photometry is performed for all sources in the image using both MAP and fixed-aperture photometry, with apertures scaled to several multiples of the point-spread function full-width half-maximum (PSF~FWHM). Light curves are constructed using the DBSCAN clustering algorithm \citep{Ester96}. Calibration of the $i^{\prime}$ and $g^{\prime}$ magnitudes is performed with an enhanced version of the ATLAS-REFCAT2 catalogue \citep{Tonry2018}, which incorporates the Pan-STARRS DR1 \citep{Chambers2016} and APASS DR10 \citep{Henden2018} catalogues. A photometric model is used to fit the light curves (in magnitudes) of matched and unmatched sources, including spatially variable zero-points, PSF-based terms, and source-specific mean magnitudes \citep{Bramich2012}. This process is done iteratively, mitigating the effects of intrinsic variability with outlier down-weighting. For SDSS $i^{\prime}$ and $g^{\prime}$ filters, XB-NEWS uses the Pan-STARRS1 standard $i_{\rm P1}$ and $g_{\rm P1}$ magnitudes in the AB system. The final model is applied to calibrate all the light curves, including those corresponding to the target. Finally, magnitudes with uncertainties $>0.25$ mag are excluded from this study, as they are considered unreliable. 

\section{Data analysis}\label{sec:data}
This section reports the procedure used to analyse the data and the main results. All the spectral results reported here have been obtained with \texttt{Xspec} \citep{Arnaud1999} version 12.14.1. In all cases, where the interstellar absorption \textsc{tbabs} model is adopted, we consistently used photoelectric cross-sections and elemental abundances from \citet{Verner1996} and \citet{Wilms2000}, respectively. 
\subsection{Multi-band light curves in the early rise phase}\label{ss:lcurves}
\begin{figure}
\centering
\includegraphics[scale=0.46]{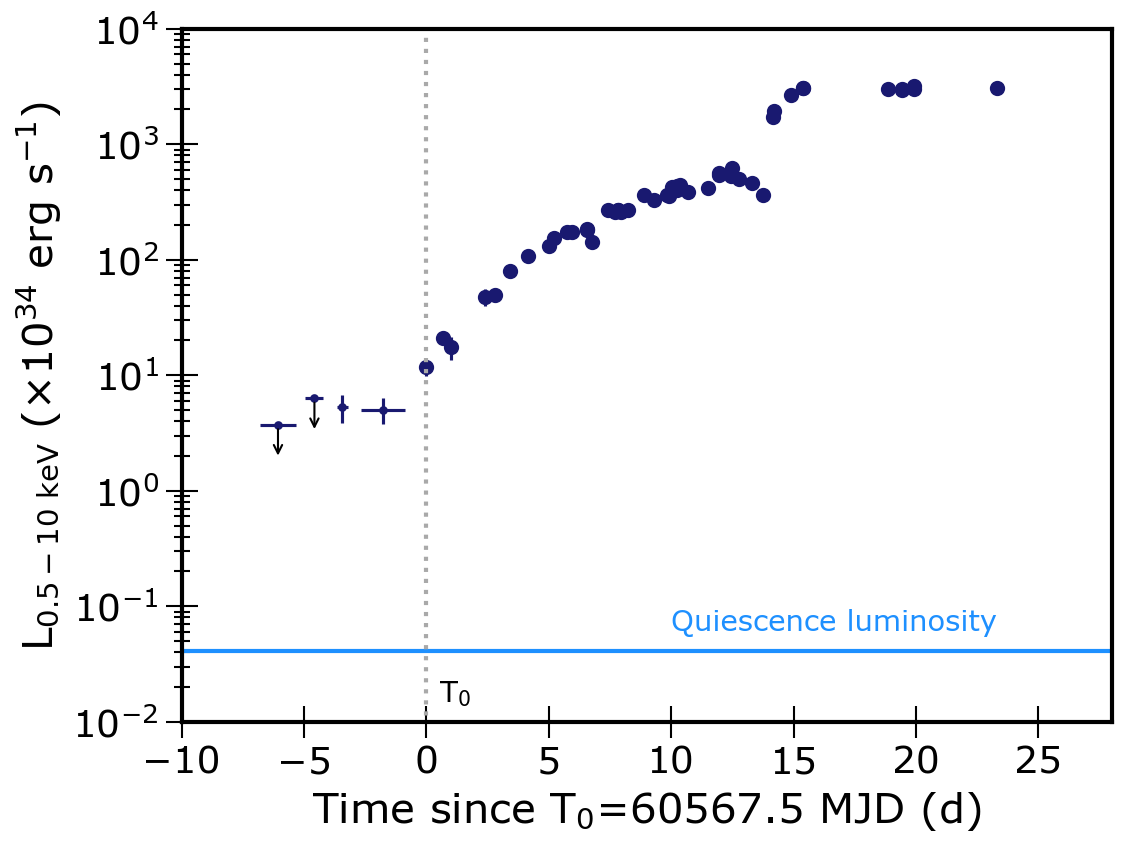}
\caption{EP/WXT light curve of the 2024 outburst of \src. The luminosity has been extrapolated to the 0.5-10 keV range (see text for more details). The quiescence level, measured by \cite{Cackett2011}, is marked with a horizontal solid light blue line, while the vertical dotted gray line indicates $T_0$, the time of the first bright (significance above 3-$\sigma$) detection by EP. The first two points correspond to non-detections and the corresponding upper limits are marked with vertical downwards arrows. The third and fourth points are instead marginal detections (significance between 2 and 3-$\sigma$) and are therefore represented with circular points but with smaller size compared to the remaining points.} 
\label{fig:ep-wxt}
\end{figure}

\begin{figure}
\centering
\includegraphics[scale=0.31]{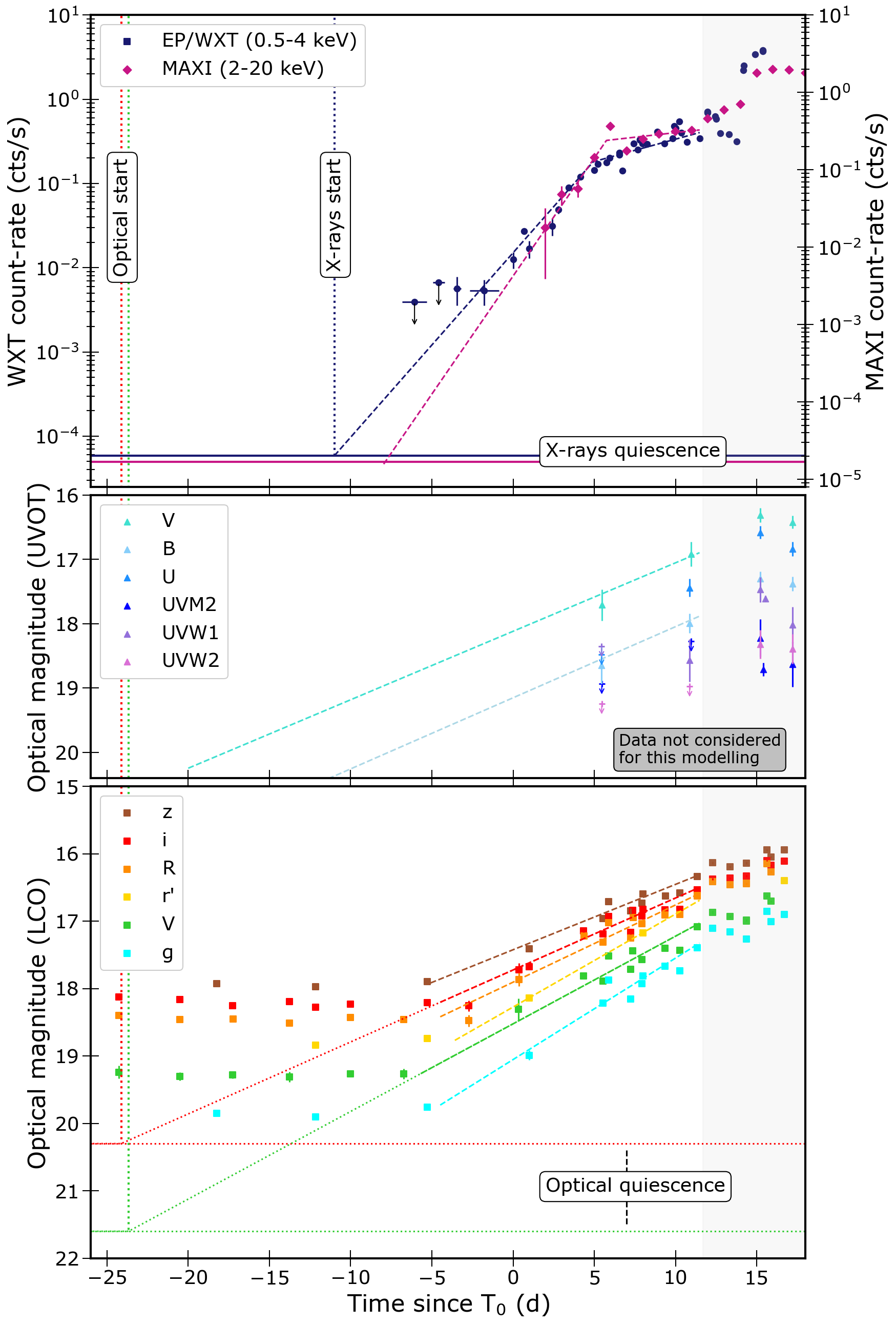}
\caption{Multi-band light curves of the outburst rise of \src, including: EP/WXT (blue) and MAXI (magenta) in the \emph{Top} panel, \swift/UVOT in different filters in the \emph{Middle} panel and LCO in the \emph{Bottom} panel. The horizontal solid lines in the \emph{Top} panels are set at the equivalent WXT and MAXI count-rates for the quiescent luminosity in \cite{Cackett2011}, while in the \emph{bottom} panel the horizontal lines indicate the real optical quiescence magnitudes measured for bands i and V by \cite{Chevalier1999}. $T_0$ corresponds to MJD 60567.5, the time of the first significant detection by EP (see text for more details). The phenomenological trends estimated for each data set are shown with superimposed dashed lines. Upper limits are marked with vertical downwards arrows.} 
\label{Fig:early-rise}
\end{figure}

Figures \ref{fig:ep-wxt} and \ref{Fig:early-rise} (\emph{top} panel) show the EP/WXT light curve of the 2024 outburst of \src in terms of X-ray luminosity and count rate, respectively. For the analysis of the EP/WXT data we systematically used an absorbed power-law model, i.e., $\texttt{tbabs}\times\texttt{powerlaw}$ in \texttt{Xspec}. Unfortunately, the poor statistics in each spectrum and the limited energy band, 0.5-4 keV, prevent us from using more sophisticated and physically reliable models. Subsequently, rather than attempting to extract physical parameters from this modeling, we use the spectral analysis solely to estimate the X-ray flux in the 0.5–10 keV range via the \texttt{cflux} component in \texttt{Xspec}. Finally, we calculate the X-ray luminosity in the above range by assuming a distance of 4.5 kpc \citep{Galloway2008}. 

In the plot, and for the rest of this manuscript, we mark the first detection of the source by EP, reported by \cite{Liu2024_atel}, as $T_0$, where $T_0=$ 60567.5 MJD. According to our results, EP first detected \src at an X-ray luminosity of about $10^{35}$ \ergs. The 4 data points prior to $T_0$ have been extracted by stacking all the WXT data taken after $T_0$-6 d by means of the \texttt{XIMAGE} software (version 4.5.1), using the \texttt{sosta} command. We used time bins varying from 2 to 0.5 days, chosen in order to accumulate at least 10 ks total exposure for point. Such a stacking procedure allowed us to obtain marginally significant (signal-to-noise ratio between 2.5 and 3.0$\sigma$) detections of the source during the 2 days before $T_0$ at a WXT count-rate (in the 0.5-4 keV band) of (5.8$\pm$1.8)$\times$10$^{-3}$ cts/s. On the other hand, even after merging the data, we could not detect the source in the previous 2-day intervals and could only derive upper limits (at confidence levels of 3-$\sigma$) of $6.7\times10^{-3}$ cts/s and $3.5\times10^{-4}$ cts/s in the $T_0 - 4 \mathrm{d}$ and $T_0 - 5  \mathrm{d}$ intervals, respectively. In order to estimate the WXT count-rate corresponding to quiescence, we used the minimum flux measured with \emph{Chandra} in quiescence as reported by \cite{Cackett2011}, 3$\times$10$^{-13}$ \flux in the 0.5-10 keV energy range, and converted it into a WXT count-rate under the assumption of an absorbed power-law spectrum with fiducial $\Gamma$ index of 2.0 and $N_H$ of 5$\times$10$^{21}$ cm$^{-2}$ (see below). We thus estimate that the X-ray flux of \src in quiescence would correspond to a WXT 0.5-4 keV count-rate of $\sim4\times10^{-5}$ cts/s. 

Besides EP, the rising phase of the \src 2024 outburst has been extensively covered by a variety of multi-band instruments. For the hard X-ray band, we collected the 2-20 keV light curve obtained by the Gas Slit Camera (GSC; 2-30 keV) onboard \maxi \citep{Matsuoka2009} from its public repository\footnote{\url{http://maxi.riken.jp/top/lc.html}}. The light curve is displayed in Fig. \ref{Fig:early-rise} \emph{Top} panel with pink diamonds. For clarity, upper limits prior to the first \maxi detection are not shown in the figure. The optical-to-UV light curves in magnitudes are displayed for all the filters used in this work in the \emph{Middle} (\swift/UVOT) and \emph{Bottom} panels (LCO) of Fig.\,\ref{Fig:early-rise}. 

The earliest UVOT observation on 2024 September 19 (MJD 60572.97) yielded only upper limits in the $u$, $uvw1$, $uvm2$, and $uvw2$ filters, while marginal detections were obtained in $v$ (17.7$\pm$0.2\,mag) and $b$ (18.7$\pm$0.3\,mag). In subsequent observations the source brightened in all filters, with the brightest measurements occurring around October 1--2 (MJD 60584--60585). The UV rise measured by UVOT showed a similar brightening to the optical light curve, confirming the rapid flux increase during the early outburst phase. 

The optical light curves are consistent with  steady flux until $\sim$MJD 60562, followed by a rising trend initially faster and then, after $\sim$MJD 60582, slower. It is noteworthy that the magnitudes measured prior to the rise by LCO (see Table \ref{tab:multiband}) do not coincide with the optical quiescence of \src; when the disc transitions to quiescence, the optical emission of \src is indeed dominated by an interloper star located $\sim$0.46$\arcsec$ from the source \citep{Chevalier1999} and by the companion star, although mostly in the K-band \citep{Matasanchez2017}. The real optical quiescence of the source has been estimated to be $\rm{mag}_{\rm V, quiescence}=21.6$ in the V-band and $\rm{mag}_{\rm i, quiescence}=20.3$ in the I-band \citep{Chevalier1999}.  Unfortunately, LCO does not have the angular resolution to disentangle the contribution from the interloper, so that in the following we will consider the LCO magnitudes prior to the outburst rise only as upper limits. 
\subsection{Light curves phenomenological analysis}
In this section, we try to estimate the time at which the outburst started ($T_{\rm start}$) at different wavebands. Such a task presents different challenges depending on the specific instrument used. Indeed, during quiescence, \src was not detected in X-rays by either EP, \maxi or \swift/BAT because too faint, and neither by LCO because contaminated by the interloper and the companion stars (see above). This lack of X-ray (optical) detection of the quiescent state makes it difficult to pinpoint the time of the X-ray (optical) rise, $T_{\rm start, X}$ ($T_{\rm start, opt}$), at which the source rose from quiescence to outburst in X-rays (optical). However, an attempt to obtain an indication of the start times at different wavelengths can be done by extrapolating the rising trend in the EP, \maxi and LCO data down to the quiescence level reported in other works. In order to do that, we used the \texttt{curve\_fit} algorithm in \texttt{scipy} to obtain a phenomenological description of the rising phase in all the light curves used in this work. Depending on the different observed trends and completeness in coverage, we used different functions:
\begin{itemize}
    \item The X-ray data, both EP and \maxi, show a general rising trend with a ``knee'' roughly 5 d after $T_0$. We thereby used a broken exponential of the form: \[
    f_X(t) =
    \begin{cases}
    A_{X}\, e^{\alpha_{1} t}, & t \leq T_{\mathrm{knee}}, \\[6pt]
    A_{X}\, e^{(\alpha_{1}-\alpha_{2})T_{\mathrm{knee}}}\, e^{\alpha_{2} t}, & x > T_{\mathrm{knee}} .
    \end{cases}
    \] where $T_{\mathrm{knee}}$ marks the time at which the index of the exponential switches from $\alpha_1$ to $\alpha_2$ and $A_X$ consists in the normalization of the exponential.
    \item The LCO data in all filters are almost flat until about 5 d prior to $T_0$ and then start rising. We therefore used a piecewise linear fit to the magnitudes (an exponential rise in flux) of the form: \[
    \mathrm{mag}_\mathrm{opt}(t) =
    \begin{cases}
    \mathrm{mag}_\mathrm{Q, apparent}, & t \leq T_{\mathrm{start, opt}}, \\[6pt]
    \mathrm{a}_{\rm opt} \,(t - T_{\mathrm{start, opt}}) +\mathrm{mag}_\mathrm{Q, apparent}, & t > T_{\mathrm{start, opt}} .
    \end{cases}
    \] where $\mathrm{mag}_\mathrm{Q,apparent}$ is the (apparent) magnitude in quiescence observed by LCO and $T_{\mathrm{start, opt}}$ is the time at which the source transitions out of quiescence in this waveband.
    \item The UVOT data are characterised by a significantly sparser coverage, preventing us from obtaining a comparably detailed description of the rise. We therefore used a simple linear function (exponential rise in flux) of the form: $\mathrm{mag}_\mathrm{UV}={\rm a}_{\rm UV}\,t+b$.
\end{itemize}
We report the results of this phenomenological analysis in Table \ref{tab:multiband}. According to such parameterization, in X-rays the rise proceeded in two steps: a steeper rise until $T_\mathrm{knee}\sim T_0+4.7$ d with index $\alpha_1=0.5$ and a slower one beyond  $T_\mathrm{knee}$ characterised by $\alpha_2\sim0.15$. The intersection between the EP quiescence level and the steeper exponential was estimated to be $T_0-11.7$ d, which can be regarded as a lower limit on $T_{\mathrm{start},X}$, assuming the rise followed the same functional form. We performed the same analysis using the \maxi data and obtained similar values for $T_\mathrm{knee}$, $\alpha_1$, $\alpha_2$ and $T_{\mathrm{start},X}$, of$\sim$ 5.6 d, 0.7, 0.17 and $>T_0-6.5$ d, respectively. 

For the optical analysis, we fixed the real quiescent magnitude level, $\mathrm{mag}_\mathrm{Q, real}$, to $\rm{mag}_{\rm V, quiescence}=21.6$ in the V-band and $\rm{mag}_{\rm i, quiescence}=20.3$ in the I-band, i.e., the values reported by \cite{Chevalier1999} without contamination from the interloper and the companion star in \src. Similarly to the procedure adopted for the X-rays light curve, we then extrapolated the linear rise observed in the V- and i- band down to $\rm{mag}_{\rm V, quiescence}$ and $\rm{mag}_{\rm I, quiescence}$ to estimate the start time of the outburst in the optical band $T_{\mathrm{start, opt}}$. For both bands, we estimate $T_{\mathrm{start, opt}}\sim T_0-24$ d. The slope of the lines shows a clear trend depending on the filter, with an increase in ${\rm a}_{\rm opt}$ with decreasing wavelength, going from ${\rm a}_{\rm opt}\sim-0.09$ with the $z'$ (IR) filter to $a_{\rm opt}\sim-0.14$ with the $g$ (blue) filter. Finally, we applied this phenomenological approach only to the data taken with filters $B$ and $U$ of UVOT, as the light curves in the other filters had too few data points. The slope a$_\mathrm{UV}$ in filter $U$, a$_\mathrm{UV}\sim-0.15$, is interestingly similar but slightly larger than the one obtained for a$_{\rm opt}$ in the blue filter. 

Along with the phenomenological analysis described in this Section, we used an alternative method to characterize the temporal evolution of the multiwavelength emission from \src\ during the early stages of its 2024 outburst within a non-parametric framework by performing Gaussian Process regression on the light curves (for a review, see \citealt{Aigrain2023}). By applying this technique, we confirm the starts of the X-rays and optical outbursts estimated with the phenomenological approach. More details about the employed methodology are provided in the Appendix, Section \ref{ss:GP}. 

\subsection{Spectral analysis}\label{ss:spectral}
\begin{figure}
\centering
\includegraphics[scale=0.40]{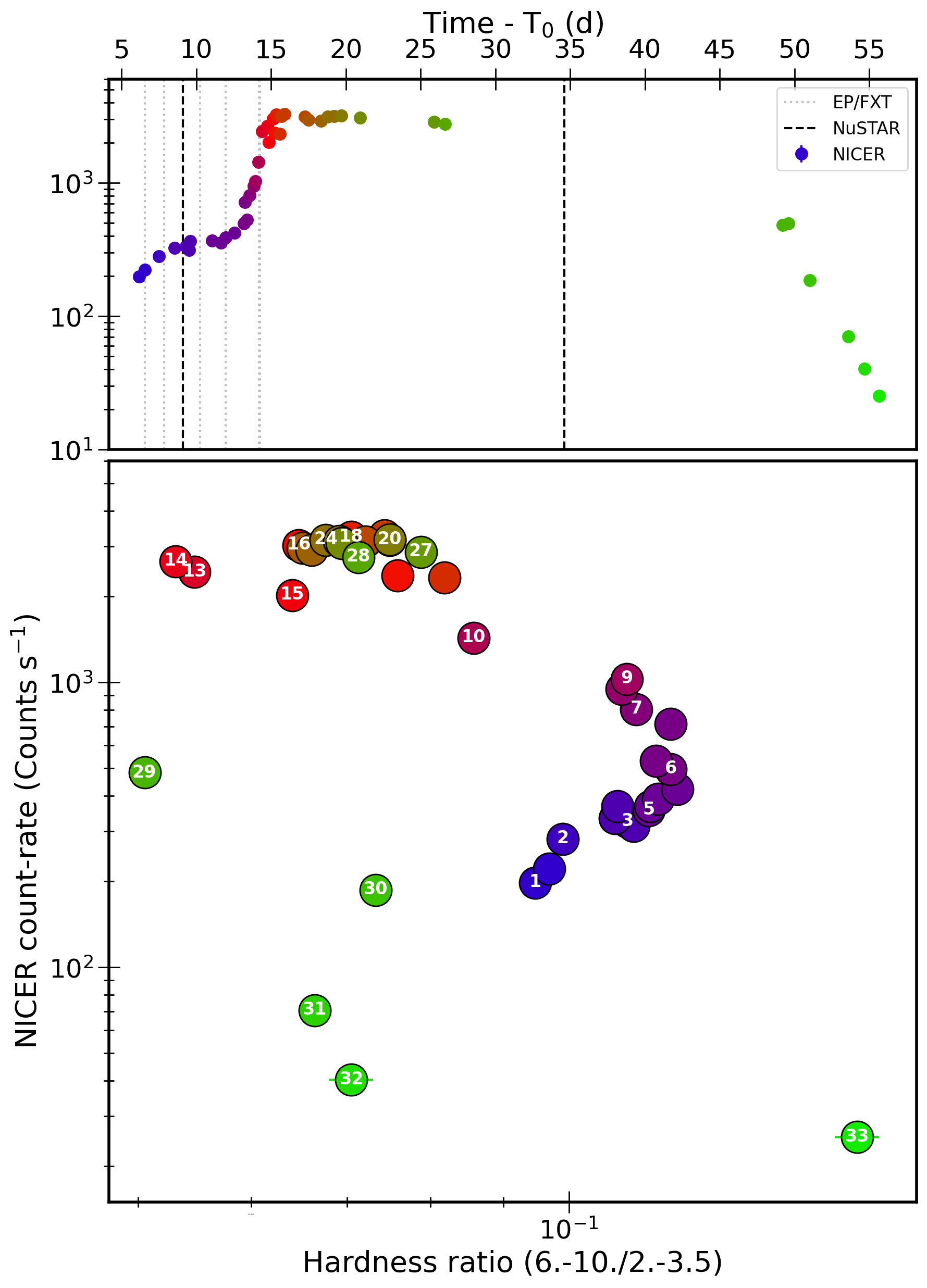}
\caption{\nicer light curve (\textit{Top}) and HID (\textit{bottom}). The total count rate has been extracted in the 1-10 keV band. The hardness is defined as the ratio between the hard band (6-10 keV) and the soft band (2-3.5 keV) count rates. A color map going from blue to red and to green is used to indicate the time evolution. The vertical lines in the \emph{top} panel mark the times of the \nustar (dashed) and EP/FXT (dotted) observations. The labels on top of selected circular data points in the \textit{bottom} panel indicate the corresponding Epoch. } 
\label{fig:lc-hid}
\end{figure}

In order to maximise the statistics and improve the quality of our spectra, we have fitted together spectra which were taken close in time and had similar spectral shape and X-ray flux. In the following, we will dub these groups of spectra ``Epochs''. Dates and corresponding spectra for each Epoch are presented in Table \ref{tab:spec-log}. In Fig. \ref{fig:lc-hid} we display the light curve and the HID produced with all the \nicer observations used in this work. In the light curve we marked down the dates of the observations with \nustar and EP/FXT. As evident from the HID, where selected points are labeled with their assigned Epoch, the outburst displayed the typical q-shaped track with a clear diagonal hard-to-soft transition from Epoch 6 on. 

Our strategy in performing the spectral analysis was to start from the two Epochs with coverage from \nustar and \nicer, Epochs 3 and 28, where the broadband spectral coverage provides reliable physical modelling. 

The spectral shapes are significantly different across the Epochs, as expected considering their distinct location in the HID, with Epoch 3 having a cut-off power-law shape extending down to the end of the nominal energy range of \nustar and Epoch 28 being characterized instead by a seemingly thermal profile ending at about 30 keV. For both Epochs, we used a blackbody component \texttt{bbodyrad} convolved with \texttt{thcomp} \citep{Zdziarski2020_thcomp}, to consider both the Comptonization spectrum and the thermal spectrum which provides the seed photons. As an extra-component in the soft X-rays was present, we included a \texttt{diskbb} component, thereby broadly consistent with the classical "Eastern" model \citep{Mitsuda1984}. The statistical need for this extra component was tested by comparing the fit with and without the additional disk contribution and then running \texttt{ftest}. We obtained negligible probabilities of improvement by chance for both Epochs, of $\sim$10$^{-61}$ and $\sim$10$^{-38}$, strongly confirming that an additional thermal component is required by the fit. When modelled with these continuum models, both spectra show clear residuals between 6-7 keV, indicating the presence of reflection. We therefore included different flavors of the \texttt{relxill} \citep[][and references therein]{Garcia2013} model to take into account reflection in different spectral states. In particular, the self-consistent \texttt{relxillCp} model \citep{Garcia2014}, which describes reflection of a Comptonized continuum off the accretion disk, was used for Epoch 3. However, as \texttt{relxillCp} assumes a seed photon temperature of 0.05 keV, which is unrealistically low for the NS LMXB case, we also multiply that component by \texttt{expabs}, a component able to mimic the low-energy roll-over expected for a higher seed photons temperature. Epoch 28 was too soft to be properly modelled with \texttt{relxillCp}, leading us to use \texttt{relxillNS} instead \citep{Garcia2022}, where the continuum incident on the disk is a blackbody. The two models used, labelled ``H'' (hard) and ``S'' (soft) for Epochs 3 and 28, respectively, are the following:

\begin{align}
\text{Model H: } \texttt{tbabs} \times ( \texttt{thComp} \times \texttt{bbodyrad} \notag \\
+ \texttt{diskbb} 
+ \texttt{expabs} \times \texttt{relxillCp} )
\end{align}

\begin{align}
\text{Model S: } \texttt{tbabs} \times ( \texttt{thComp} \times \texttt{bbodyrad} \notag \\
+ \texttt{diskbb} + \texttt{relxillNS} )
\end{align}

\begin{figure*}
\centering
\includegraphics[scale=0.25]{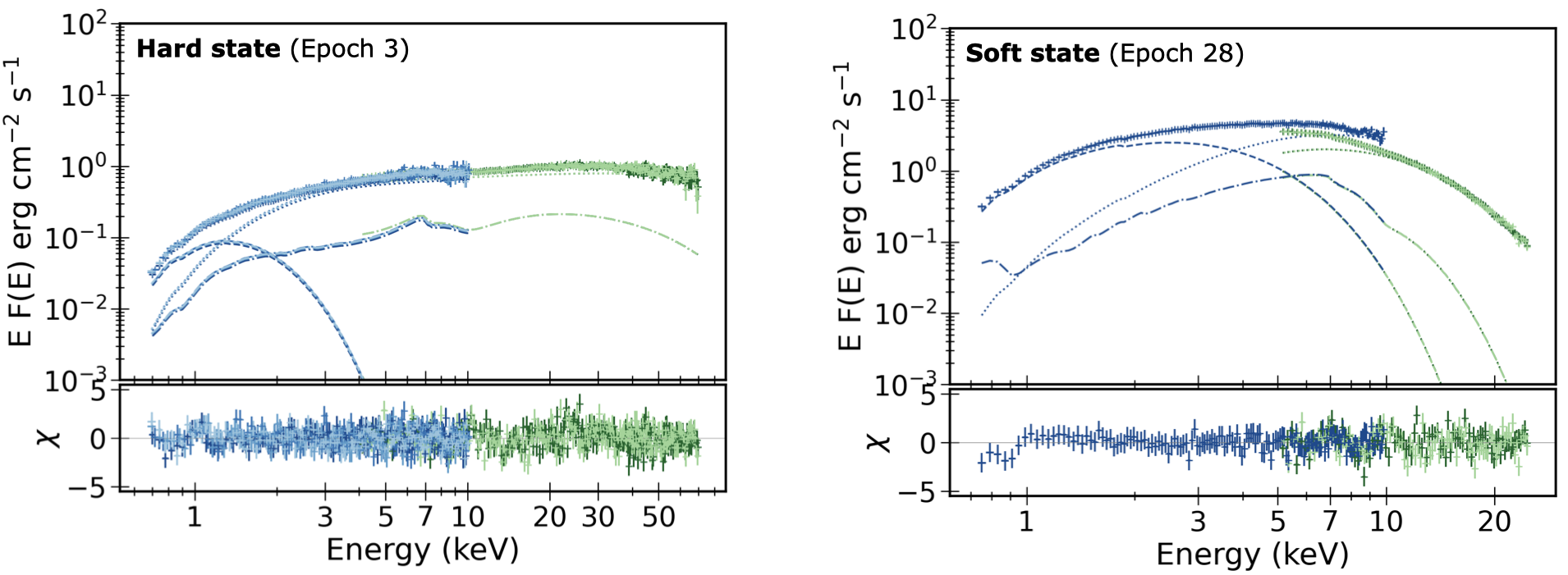}
\caption{Broadband \nicer (different shades of blue to distinguish between the \nicer spectra used in the same Epoch) and \nustar (different shades of green to distinguish between FPMA and FPMB) spectra for Epoch 3 (\textit{left}, fitted with Model H) and Epoch 28 (\textit{right}, fitted with Model S) and residuals. Different line styles were adopted to distinguish between the different components: dash for \texttt{diskbb}, dot for \texttt{thcomp}$\times$\texttt{bbodyrad} and dash-dot for \textsc{relxillCp} (\textit{left} panel) and \textsc{relxillNS} (\textit{right} panel).} 
\label{fig:broadband}
\end{figure*}

An overview of all the parameters included in the models and their best-fit values are reported in Table \ref{tab:fit_broadband}. In Model ``H'', for physical consistency, we tied $E_{\rm cut-off, low}$ to $3\times kT_{\rm bb}$, we bound together the photon indices $\Gamma$ and electron temperatures $kT_{\rm e}$ of \texttt{thcomp} and \texttt{relxillCp} and fixed the reflection fraction parameter $f_{\rm refl}$ to -1 so that we only consider the reflection spectral component. Some parameters were not well constrained by the fit, so that we froze them to typically adopted values for LMXBs \citep[e.g.][]{Saavedra2023, Ludlam2024, LaMonaca2025}: the emissivity index $\epsilon$ was set to 3, the Fe abundance A$_{\rm Fe}$ to 1 and the density parameter $\log{N}$ to 19. We show the broadband spectra for Epochs 3 and 28, along with the best-fit models used to describe them, in Fig. \ref{fig:broadband}. 

For the remaining Epochs, with the exception of Epochs 30-34 (see below), we have adopted Model ``H'' or Model ``S'', but under some assumptions, as the lack of data above 10 keV limits significantly the quality of our fits. First of all, we fixed $kT_e$ in \texttt{thcomp} to 23.0 keV (3.5 keV) for the spectra modelled with Model ``H'' (``S''), i.e. to the value obtained for Epoch 3 (28), which was taken in a similar state. While realistically we expect $kT_e$ to change throughout the outburst, we consider these assumptions acceptable as: i) for spectra described with ``H'', where $kT_e$ is above 10 keV, any impact that variations of $kT_e$ may have on data with only soft X-ray coverage would be negligible; ii) for spectra described with ``S'', the limited statistics does not allow to constrain $kT_e$, so that the fit is mostly insensitive to variations of this parameter. Additionally, for similar reasons, we froze all the parameters of the reflection component, being it \texttt{relxillCp} or \texttt{relxillNS}, with the exception of the normalisation $K_{\rm refl}$. While this general strategy allows us to obtain acceptable fits for each Epoch, we of course will not consider this as a proof that the reflection parameters, as well as $kT_e$, do not change from one Epoch to the other. Finally, a different strategy has been adopted for the last five Epochs, corresponding to the decay phase of the outburst. Given the much lower count-rate compared to the rest of the outburst, we had to first remove the reflection component (Epochs 29--31) and then also the disk component (Epochs 32--33), as they were no longer required by the fit. 

\begin{figure}
\centering
\includegraphics[scale=0.35]{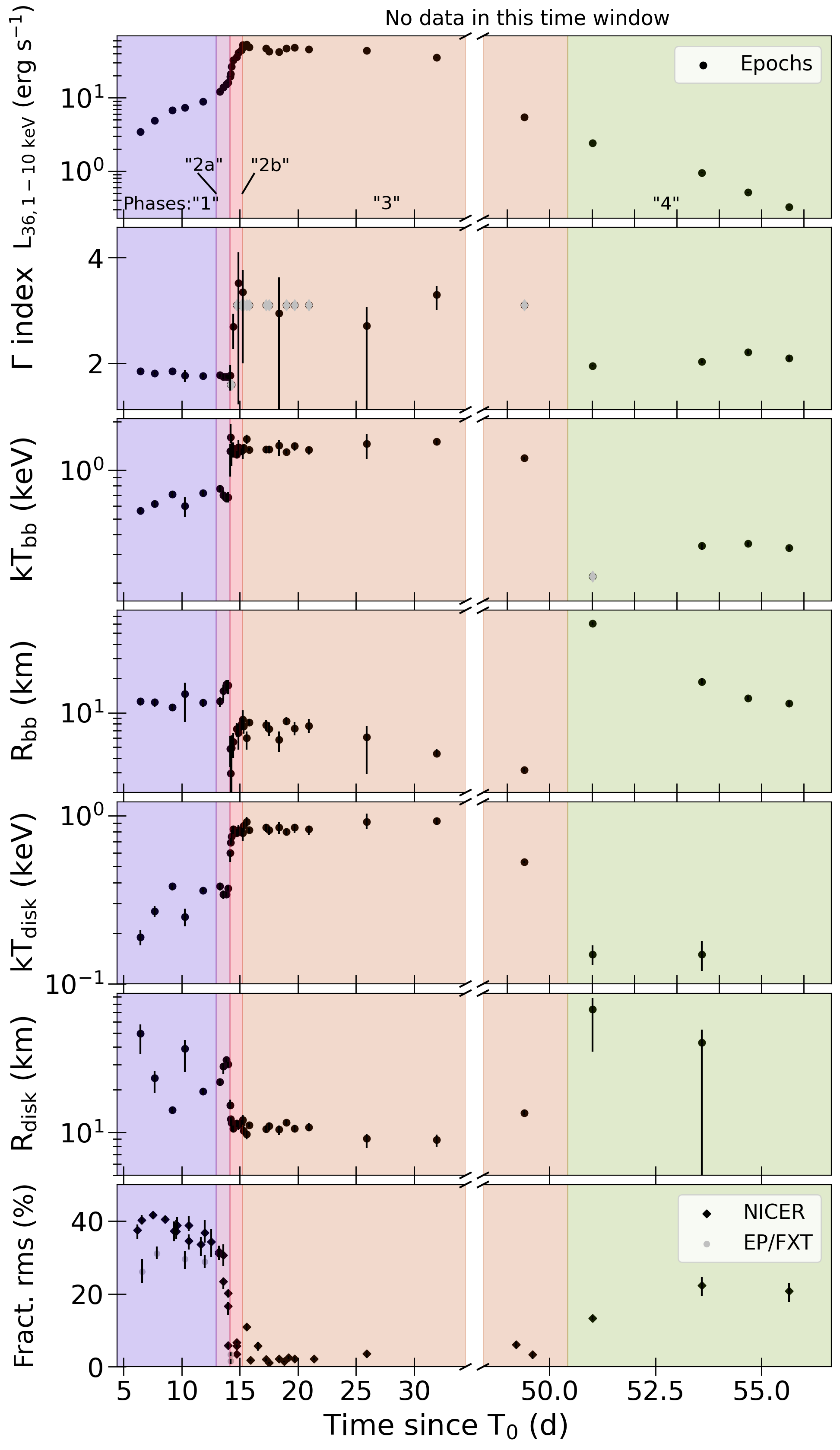}
\caption{Evolution of the main spectral-timing parameters of \src during the 2024 outburst. In the top six panels (Luminosity in the 1 - 10 keV band, $\Gamma$ index of the comptonizing medium, temperature and radius of the black body component, temperature and radius of the disk component), each point corresponds to a different Epoch. In the bottom panel (fractional rms), we distinguish instead between values obtained for \nicer (black diamonds) and EP/FXT (gray circles). The fractional rms has been calculated considering the 0.01-1000 Hz frequency range and the 1-10 keV energy range. Parameters that were kept frozen during the fit are reported as gray diamonds.} 
\label{fig:tower-plot}
\end{figure}

The results of our individual fits are presented in Table \ref{tab:individual-fits} and in the tower plot in Fig. \ref{fig:tower-plot}. From the normalization values of the \texttt{bbodyrad} ($K_{\rm bb}$)  we extracted the radius of the blackbody emitting region $R_{\rm bb}$ (in km) using the formula: $K_{\rm bb}=(R_{\rm bb}/D_{\rm 10})^2$, where $D_{\rm 10}$ is the source distance in units of 10 kpc, set in this case to 0.45 \citep[as reported by][]{Galloway2008}. Similarly, we estimated the disk inner radius from the normalization of \texttt{diskbb} ($K_{\rm disk}$) using the formula: $K_{\rm disk}=(\xi f^2_{\rm color})^{-2}(R_{\rm disk}/D_{\rm 10})^2 \cos{i}$, where $\xi$ is the \cite{Kubota1998} correction factor equal to roughly 0.4, $f_{\rm color}$ is the color correction factor equal to 1.7 \citep{Shimura1995} and $i$ is the system inclination, fixed to 30$^\circ$ according to the results of this and previous papers \citep[e.g.][]{Ludlam2016}. We note that the correction factor $\xi$ accounts for the torque-free condition, which might not be always valid for NS systems, so that our estimates for $R_{\rm disk}$ are probably systematically underestimated and should not be taken at face value. 

We also tested an alternative spectral configuration where the \texttt{bbodyrad} component, accounting for the blackbody emission arising from the NS/boundary layer, is the softer component, while the seed photon spectrum for the Comptonization component is modelled with \texttt{diskbb}. In this scenario, which resembles the classical ``Western'' model \citep{White1988}, the blackbody emission from the NS/boundary layer is observed directly, while the disk emission is entirely Compton-scattered by the corona. Although the resulting fits are statistically acceptable and comparable to those obtained with models H and S, they return a similar configuration to our original ``H'' and ``S'' models with just \texttt{bbodyrad} and \texttt{diskbb} interchanged, i.e., where the radius of the soft component observed directly is larger than the radius of the component observed through Compton-scattering. In other words, using a "Western" model systematically yields $R_{\rm disk}$ smaller than $R_{\rm bb}$. Furthermore, in the soft-state spectra, $R_{\rm disk}$ falls below $\sim$5 km, which, even allowing for potential underestimation (see above), is physically implausible, as it would imply emission originating within the NS radius. For the rest of the paper, we will therefore only discuss the results obtained with the original (``Eastern model''-like) models ``H'' and ``S''. 

\begin{figure}
\centering
\includegraphics[scale=0.35]{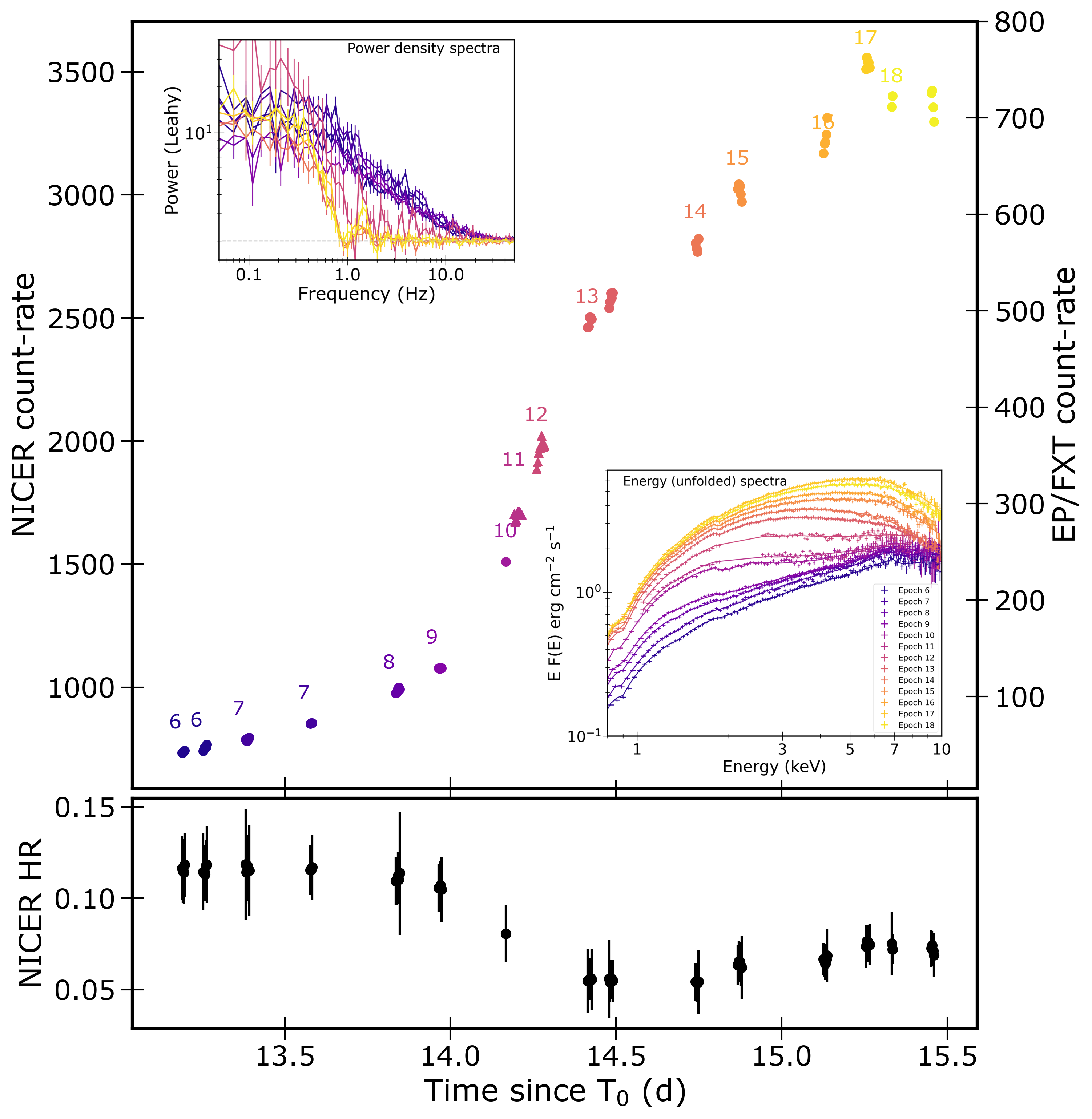}
\caption{\nicer and EP/FXT light curves (\emph{Top}) and hardness ratio (\emph{Bottom}) during the spectral state transition period occurred between MJD 60580 and 60583. In the \emph{top} panel, circles are used for the \nicer observations and triangles for the EP observations. A time-bin of 240 s has been used for each observation. The Epoch in which each observation has been classified is indicated with superimposed labels. The inset plots show the evolution of the power density (top-left inset) and energy (bottom-right) spectra for the period considered.} 
\label{fig:transition}
\end{figure}

As evident from the tower plot, the outburst can be broken down in 4 phases (highlighted with different colors in the plot): a relatively ``slow'' rise lasting for a week (Phase ``1''), a dramatically rapid, a few days-long, flux increase (``2a'', ``2b''), a several week-long plateau at the outburst peak (``3'') and finally a decay to quiescence (``4''). The highly dynamic spectral variability in Phases ``2a'-``2b''  can also be inferred by the fact that all the Epochs from 6 to 20 have been taken between September 27th to September 30th. In order to zoom-in on this particularly eventful phase of the outburst, we show a 3-day light curve of this time period, along with the different energy spectra, power density spectra (see Section \ref{ss:timing}) and hardness ratio in Fig. \ref{fig:transition}. The plot clearly shows that in about half a day the count-rate more than doubles, the hardness ratio drops by a factor of 3, and the shapes of both the energy and power density spectra change abruptly and dramatically. 

\subsection{Timing analysis}\label{ss:timing}
For the timing analysis, we corrected the photon arrival times of the source event files to the Solar System barycentre using the \texttt{barycorr} tool, the most recent calibration files, and the DE405 ephemeris. We adopted the known coordinates of the source, R.A. = 19$^\mathrm{h}$11$^\mathrm{m}$16$\fs$05713, Decl. = +00$^{\circ}$35$^{\prime}$05$\farcs$8682 (J2000.0) \citep{Gaia2020}. We considered all the X-ray observations used in this work, selecting as energy ranges: 0.3--12\,keV for \nicer and EP and 3--79\,keV for \nustar. Single observations split into ``Epochs'' for the spectral analysis were here analysed together. We extracted the power density spectrum (PDS) by averaging over 64-s long segments with time bins of 1.0\,ms. In order to correct for the dead-time, we applied the \nustar PDS using the Fourier Amplitude Difference (FAD) method \citep{bachetti2018} as implemented in {\tt Stingray} \citep{Huppenkothen2019,Bachetti2024}. For each PDS, we estimated the fractional rms over the entire frequency band by modelling the PDS in {\sc Xspec} with a combination of Lorentzian functions and a constant to account for the Poisson noise contribution. We show the temporal evolution of the rms for the 0.3--12\,keV \nicer PDS in the bottom panel of Figure\,\ref{fig:tower-plot}. 

In order to estimate the ``break'' frequency $\nu_{\rm break}$ dividing the band-limited noise dominated part of the PDS from the red noise \citep{Wijnands1999}, we also analysed each PDS with a single Lorentzian with centroid fixed at 0 and used the estimated width as a proxy for $\nu_{\rm break}$ (see Fig. \ref{fig:nu_break}). We systematically inspected all PDS searching for QPOs and/or other discrete features over the 0.01-1000 Hz frequency range. In order to do that, we searched whether our fit with a sum of broad Lorentzians left any fine structures unmodelled in the residuals and modelled such features with an extra Lorentzian. We then considered significant all the features for which the ratio between the Lorentzian normalization and its 1-$\sigma$ error was found higher than 3, corresponding to a statistical significance of 3-$\sigma$. No significant discrete component was found from this search.

\section{Discussion}\label{sec:disc}

\subsection{The ``dawn'' of the outburst}
The early phases of LMXB outbursts represent, to date, a significantly underexplored regime of the activity from these sources. In this work, we have presented a multi-band monitoring campaign of the first $\sim$ two weeks of the \src 2024 outburst from optical (LCO) to UV (UVOT) up to soft (EP/WXT) and hard (\maxi) X-rays. 

Using both a phenomenological approach and Gaussian Processes, we have estimated that: i) assuming that the linear rise in magnitude in the I- and V-bands can be extended down to the quiescent level, we can propose a lower limit on the beginning of the optical outburst to MJD 60543, $\sim 24 \ $d prior to the first EP detection, $T_0$ \citep{Liu2024_atel}; ii) by doing an analogous assumption on the trend observed in the EP data, the beginning of the X-ray outburst had to start after MJD 60556, $\sim 11 \ $d prior to $T_0$; iii) \src was detected by EP in the two days prior to $T_0$ (albeit with a low SNR of 2.5-3), on MJD 60565, suggesting that the source was already active at that stage. These considerations suggest that the optical brightening preceded the onset of the X-ray emission, implying a possible optical-to-X-ray delay of $\lesssim$ 13 days. This is expected according to the standard Disk Instability Model \citep{Lasota2001}, where the initial rise is driven by the formation of a heating front that propagates through the accretion disk, increasing the optical contribution from the disk as larger portions are progressively heated. During the early phases, however, X-ray production is expected to be dominated by Compton up-scattering of disk photons in a hot inner flow, which becomes efficient only after the heating front reaches the inner regions of the disk. Detectable X-ray emission would therefore emerge only after the optical rise, naturally accounting for the inferred delay. Furthermore, our result is in line with all the LMXBs for which a delay between the optical and X-ray onsets has been measured \citep[see][and references therein]{Rout2025_1807}, where the optical rise is observed to precede the X-ray rise. Regarding NS LMXBs in particular, our measured optical-to-X-ray delay is larger than the one obtained for the 1997 outburst of \src \citep[3-d,][]{Shahbaz1998}\footnote{However, this delay must be interpreted with caution. First, it was measured with the All-Sky Monitor onboard the Rossi X-ray Timing Explorer, whose sensitivity was more than an order of magnitude lower than that of EP/WXT \citep{Yuan2022}. Second, the optical magnitude assumed to represent the quiescent level was likely overestimated, as it included contamination from the interloper star.}, for SAX J1808.4-3658 \citep[4 d,][]{Goodwin2020} and MAXI J1807+132 \citep[4--12 d,][]{Rout2025_1807}, which are to date the only other NS LMXBs for which optical-to-X-ray delays have been reported.

Finally, we stress that our measured optical-to-X-ray delay depends critically on the assumption that the trends observed in X-rays and optical during the early rise can be extended down to quiescence. A different delay, and even a scenario where X-rays actually preceded the optical emission, although unlikely, can not be fully excluded.

\subsection{The hard state (Phase ``1'')}\label{ss:hard-state}
Our detailed spectral and timing monitoring of the source started 6 days after the first EP detection of the source and it consisted of 33 ``Epochs'' throughout the entire outburst. All spectra from Epoch 1 to 31 were analyzed with a model consisting of a single Compton-scattered blackbody emission, disk blackbody emission, and a reflection component (models ``H'' and ``S''). For the latter component, two different models were applied depending on the spectral shape of the incident spectrum: during Epochs 1–9 and 30–31, a Comptonization model (\texttt{relxillCp}) was used, whereas for Epochs 10–29, the incident spectrum was assumed to be a blackbody (\texttt{relxillNS}). According to the results of our spectral analysis and the main physical parameters trends displayed in Figures \ref{fig:tower-plot} and \ref{fig:nu_break}, we can reconstruct the evolution of \src during its 2024 outburst in five phases: ``1'', ``2a'', ``2b'', ``3'' and ``4''. Phase ``1'' lasted for about 7 days and entails Epochs from 1 to 5. 

Throughout this phase, the source shows a hard spectral shape, with a photon index $\Gamma$ of $\sim$1.8. The blackbody temperature and radius are well constrained at around $\sim0.7$ keV and $\sim$10-12 km, respectively. The size of the blackbody component suggests that it arises from about the entire NS surface. On the other hand, the disk is significantly colder than the black body component, but it gradually heats up throughout this phase, going from an initial temperature of $0.17$ keV to $0.36$ keV at the end of this phase. At the same time, the inner radius of the disk, estimated from the normalisation of \texttt{diskbb}, shows an initial decreasing trend from about 100 km to about 20 km, and then increases again, starting a puzzling trend that will be discussed further in the next section. The temperature of the electron corona $kT_{\rm e}$ can only be constrained during Epoch 3, the only one in this phase with \nustar\ coverage, giving a value of $\sim$22 keV, a standard value for atolls in such a state \citep[e.g.][]{Burke2017,DiSalvo2019,Anitra2021,Marino2022,Banerjee2024,Illiano2024}. The analysis of the reflection component for the same epoch provided a measurement of the system inclination, in the range 25$^\circ$-33$^\circ$, compatible with what was found by \cite{Ludlam2016} for the same source, and further confirming \src as a moderately low-inclination LMXB. The inner disk radius can be constrained from the analysis of the reflection component in Epoch 3, where an upper limit of 10 gravitational radii R$_{\rm G}$, about 20 km for a 1.4 M$_\odot$ NS, was obtained, which is consistent with the disk radius measured from the \texttt{diskbb} normalization. Throughout this stage the fractional rms was relatively high, but it showed a general decreasing trend, going from 40\% to about 30\%. On the other hand, the value of the PDS break frequency $\nu_{\rm break}$ showed a consistently increasing trend (see Fig. \ref{fig:nu_break}), which is considered a signature of the accretion disk approaching the compact object \citep[e.g.][]{Churazov2001}. 

\begin{figure}
\centering
\includegraphics[scale=0.29]{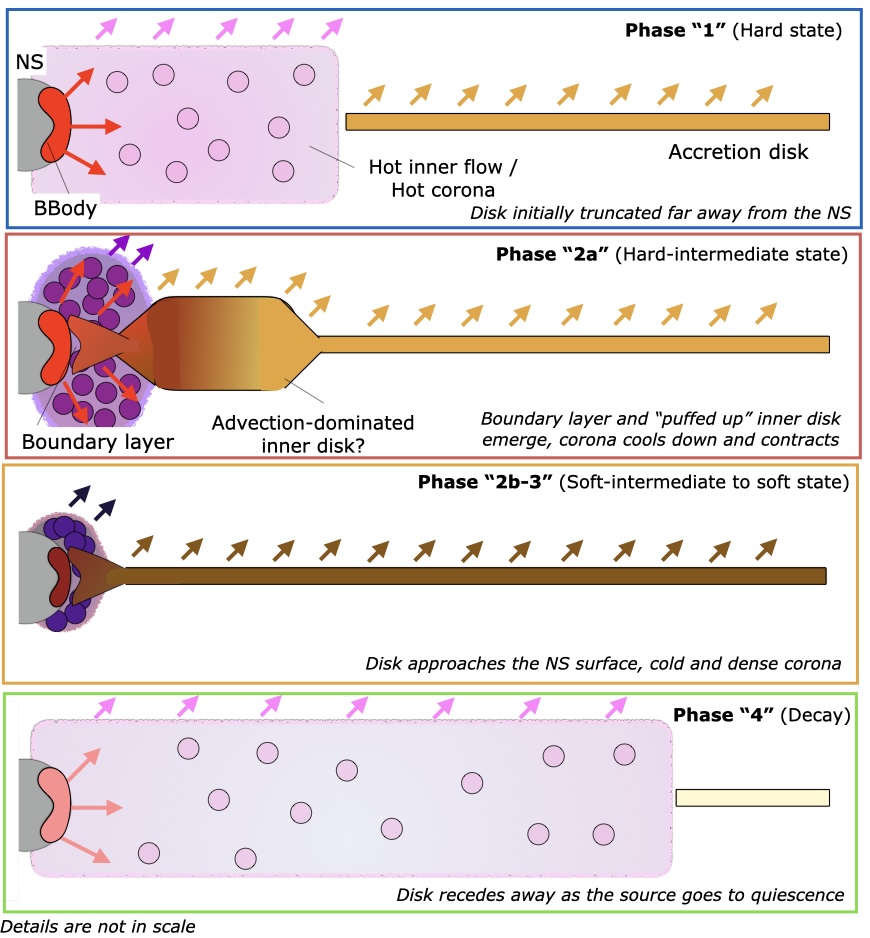}
\caption{Toy model of our proposed evolutionary scenario for the 2024 outburst of \src. The intensity of the color is qualitatively proportional to the temperature of each region; in particular the temperature increases from purple to pink for the hot inner flow, from yellow to brown for the disk and from salmon to dark red for the blackbody emission. The number of circular points inside the hot flow sketch represent its density.} 
\label{fig:toy-model}
\end{figure}

The spectral and timing parameters are consistent with the accretion disk being truncated away from the NS and a rather hot inner flow or corona Compton-upscattering photons from the NS surface (as we sketch in the top panel of Fig. \ref{fig:toy-model}), a scenario consistent with a canonical hard state.

\subsection{From hard to intermediate: witnessing the emergence of spreading and boundary layers? (Phase ``2a'')}\label{ss:boundary-layer-a}
We classified in Phase ``2'' all the Epochs from 6 to 16. Despite the number of epochs being greater in number than the previous phase, the entire Phase ``2'' is remarkably shorter, lasting about 2 days. During this short time window, the source displays a dramatic evolution, which we break down here in two different sub-phases (``2a'', ``2b''). 

Initially, while $\Gamma$ remains at the same value of Phase ``1'', the fractional rms starts to drop significantly from $\sim$30\% to $\sim$20\% and the luminosity starts to increase significantly, going from $\sim$7.2$\times$10$^{36}$ \ergs to 1.2$\times$10$^{37}$ \ergs in about one day. In our fits, the increase in luminosity manifests as an increase in the normalization, and therefore the size of the emitting region, of both the blackbody component and the disk component. Such a trend is statistically significant and is unlikely to be driven by parameter degeneracies, as indicated by the absence of secondary minima in the $K_{\rm bb}$–$K_{\rm disk}$ contour plots for two representative epochs of Phases “1” and “2a” (see Fig.~\ref{fig:contour-plots}).

The most trivial way to interpret the trend in the disk component would be to invoke a receding of the disk away from the compact object. However, if the disk was actually moving away, we would expect both its temperature $kT_{\rm disk}$ and the PDS break frequency $\nu_{\rm break}$ to decrease; either trend is visibly absent from our results (see Fig. \ref{fig:tower-plot} and \ref{fig:nu_break}). Moreover, a scenario in which the disk recedes during the state transition would be inconsistent with the canonical truncated-disk model \citep[e.g.][]{Done2007}. In this framework, NS LMXBs in the hard state are characterized by a vertically extended inner hot flow replacing the innermost regions of the disk \citep[e.g.][]{Narayan1995,Done2007,PoutanenVeledina14}. As the mass accretion rate and luminosity increase, the disk is instead expected to move inward toward the compact object. By when the source reaches the soft state the disk should be a thin disk \citep{ShakuraSunyaev1973} with a low contribution from a significantly cooled down inner flow/corona. This behavior is indeed what is typically observed in NS LMXBs undergoing state transitions \citep[e.g.][]{Marino2019b, Rout2025_1807_spectral}. When exactly the transition happens is still under debate \citep[see e.g. discussions in][]{Poutanen18,Zdziarski21a,Zdziarski21b}. In the case of NSs, the accretion flow needs also to slow down from the quasi-keplerian azymuthal velocity profile of the disk to match the spin of the star (this transition can take the form of a BL/spreading layer, depending on the mass accretion rate; e.g. \citealt{Popham2001,Suleimanov2006, Done2007,Inogamov1999,InogamovSunyaev2010}). 

In order to resolve the tension between our data and the above model, we discuss two possible scenarios where an increase in disk normalization can be compatible with a constant (or even decreasing) inner disk radius. An increase in the disk normalization, which is a rough measure of the number of photons emitted by the region, could be explained with a sudden geometrical thickening, or a ``puffing up'', of the disk as it pushes inside to approach the NS surface\footnote{ We caution however, that this effect may be mitigated by the moderately low, i.e., $\sim$30$^\circ$, inclination of \src (see Section \ref{ss:hard-state}).}. Indeed, at sufficiently high accretion rates \citep[but still sub-Eddington, e.g.][]{Straub2011}, part of the energy might be removed by advection instead of being radiated away, as in the ``slim disk'' model \citep{Abramowicz1988}. It is noteworthy that the simultaneous contraction and cooling of the corona during Phase ``2a'' may also contribute to the increase in the disk scale height. In contrast, a hotter corona would maintain the outer disk layers highly ionized, reducing the optical depth and thus inhibiting the formation of a slim disk \citep[e.g.,][]{Peng2024}.

Since this regime has different properties than the thin disc, we have mimicked this scenario by replacing \texttt{diskbb} with \texttt{diskpbb} \citep{Mineshige1994,Kubota2005}, taking into account that turning up advection causes a deviation in the radial dependency of the disk temperature, $T\propto R^{-p}$, from a \cite{ShakuraSunyaev1973} thin disk. This model is routinely used for Ultraluminous X-ray sources \citep[e.g.][]{Bachetti2013,Pintore2025} and while the \src accretion rate is by no means comparable to those sources, we still suggest that the structural modifications experienced by the accretion flow during the transition observed in Phase ``2a'' may be analogous. We show in Appendix \ref{ss:app-slim-disk} that using \texttt{diskpbb} both the increases in the inner radius of the disk $R_{\rm disk}$ and in size of the black body component $R_{\rm bb}$ can be replaced by a slight decrease of $p$, going from $\sim$0.75 (as expected for Shakura \& Sunayev thin disks) to about 0.55. The scenario of a disk ``puffed up'' by advection may suggest that during Phase ``2a'' we are witnessing the emergence from the shrinking corona of the spectrum from such inner disk, which is expected to be more advection-dominated, and therefore more vertically extended, with respect to a \cite{ShakuraSunyaev1973} thin disk. We sketch our proposed scenario in Fig. \ref{fig:toy-model}.

An alternative possibility is that the trend in $K_{\rm disk}$ may be driven by changes in the correction factors $f_{\rm color}$ and/or $\xi$, which have been assumed constant throughout the outburst. For instance, $f_{\rm color}$ may have been higher in the hard state, as discussed by \citet[][and ref. therein]{Ren2022} for a different LMXB, and reached its canonical value of 1.7 only during the state transition. 

\subsection{A 12-h state transition? (Phase ``2b'')}\label{ss:boundary-layer-b}
Going from Epoch 9 to Epoch 12, the spectral shape of the source (see inset in the low-right corner of Fig. \ref{fig:transition}) changed dramatically within about 12 hours. The evolution observed between these epochs is arguably one of the best documented spectral transitions in a LMXB to date, with almost hourly coverage thanks to the combined observations of \nicer and EP. Because of the transition, we changed the best-fit model from model ``H'' to model ``S'' going from Epoch 9 to Epoch 10, meaning that the spectral shape of the spectrum illuminating the reflecting disk changed from a cut-off power-law shape to a blackbody. 

$\Gamma$, $kT_{\rm bb}$ and $kT_{\rm disk}$ all rapidly increase to values of about $\sim$3, $\sim$1.4 and $\sim$0.7 keV. An even more dramatic change can be observed for the physical sizes of both the black body component and the disk; as the latter suddenly approaches the NS surface, with the (apparent) inner radius moving from $\sim$40 to $\sim$15 km within a few hours, the size of the black body shrinks by a factor 4, to a size of 3--4 km, narrow region on the NS surface, at the boundary between the disk and the NS. The fractional rms continues its rapid decrease, dropping to a level of only 3--6\% along the transition. While we cannot constrain the electron temperature in this phase, the rms evolution and the overall spectral softening strongly suggest that the corona may have experienced significant cooling during this epoch. We also note that as the corona becomes cold and dense, it will cover a smaller fraction of the NS surface (see Fig. \ref{fig:toy-model}, c), explaining the sudden shrinking of blackbody size $R_{\rm bb}$ in Phase ``2b'', possibly corresponding to the boundary layer becoming visible \citep{Inogamov1999,InogamovSunyaev2010} and indicating the imminent
soft state. Interestingly, such a trend during state transitions has been also invoked by \cite{Cavecchi2026} in a different context, where it was used to explain why Type-I X-ray bursts have different recurrence times trends between NS LMXBs spectral states. 

The ensemble of these spectral and timing clues clearly indicates that Phase 2 coincides with a remarkably rapid epoch of hard-to-soft transition and it can be identified as the intermediate state of \src. The notion that \src experiences rapid, of the order of days, state transitions has been suggested before. \cite{MunozDarias2014} report in their global study of hysteresis loops in NS LMXBs a hard-to-soft time scale of 1-4 days for \src, a result which can be also confirmed by visually inspecting light curves and/or hardness intensity diagrams of previous outbursts from the source \citep[e.g.][]{Tudose2009,DiazTrigo2018} and by the recent broadband study of the 2023 outburst of the source by \cite{Yan2025_aqlx1}. In the 2012 outburst of the transient NS LMXB 1RXS J180408.4-365028, a short-lived intermediate state was also identified and a hard-to-soft transition time-scale of 1 d suggested \citep{Marino2019b, Fiocchi2019}. Our results, however, allow for the first time to precisely identify the timescale of the full transition and to have unprecedented insights on the physical mechanisms underlying spectral transitions in NS LMXBs. A causal connection between the increase in mass-accretion rate, the increasing radiation from both the boundary/spreading layers and the disk and the subsequent cooling/disruption of the corona could explain such a rapid state transition in \src. Beyond this single source, we propose that similar mechanisms may be at play in all state transitions in NS LMXBs. Indeed, while NS LMXBs typically perform state transitions in only a few days (see references above), LMXBs with BHs are often characterised by much longer hard-to-soft transitions, of dozens to hundreds of days \citep[][but see \citealt{Bock2011} for a reported very rapid state transition in the BH X-ray binary Cyg X-1]{Tetarenko2016review}. Because accreting BHs lack the additional emission from the NS surface and/or BL, the accretion disk is the sole source of soft photons available for coronal cooling in these systems, perhaps making the state-transition timescale significantly longer than in NS systems.

\subsection{The soft state (Phase ``3'')}\label{ss:soft-state}
After the spectral turmoil in Phase ``2'', Phase ``3'' consists in a much longer and stable phase for the source. All spectra in this phase show consistent values in $\Gamma$ ($\sim$3.0), $kT_{\rm bb}$ ($\sim$1.3 keV), $kT_{\rm disk}$ ($\sim$1.0 keV), $R_{\rm bb}$ ($\sim$8 km), $R_{\rm disk}$ ($\sim$15 km) and fractional rms well below 10\%. Thanks to the broadband spectral analysis of Epoch 28, we can constrain the value of $kT_e$ in this phase to $\sim$3.3 keV. In this Epoch, the values of the inner disk radius provided by both \texttt{diskbb} and the reflection component \texttt{relxillNS} are well-constrained but only marginally consistent with each other, as the reflection component points to a slightly larger disk. We stress, however, that our estimates for $R_{\rm disk}$ are likely underestimated and, while the observed trends in these parameters are reliable, they should not be taken as real values. The $R_{\rm in}$ found by \texttt{relxillNS} is also compatible with the values obtained by the broadband spectral analysis of \src for two previous outbursts \citep{Ludlam2017}. The inner radius is not the only physical parameter that seems to be the same between the broadband spectra in Epochs 3 and 28. The disk ionization  $\log{\xi}$ and the reflection normalization $K_{\rm refl}$ seem to be broadly consistent as well between the Epochs, suggesting that the spectral evolution from one state to the other only marginally affects the reflection parameters and therefore the reflecting medium. 

It is noteworthy that while all the values above are also consistent with the parameters achieved at the end of Phase ``2'', the size of the blackbody radius shows an increase with respect to the previous Phase. A plausible interpretation of this trend is that as it is being fed by the accreted material, the size of the spreading layer increases throughout this phase, heating up a gradually larger fraction of the NS surface with respect to Phase ``2'' \citep{Inogamov1999,InogamovSunyaev2010,Suleimanov2006}. We conclude that throughout Phase ``3'', the source lingered in a long and almost uneventful soft state.

\subsection{The ``dusk'' of the outburst (Phase ``4'')}\label{ss:decay}
In the final phase, which encompasses Epochs 30 to 33, the X-ray luminosity falls down below $10^{36}$ \ergs as the source decays towards quiescence. The spectrum hardens significantly, with $\Gamma$, $kT_{\rm bb}$ and $R_{\rm bb}$ and the fractional rms returning to the values observed in Phase ``1''. However, Phase ``4'' cannot be considered as a carbon copy of Phase ``1'', mostly because of a much lower contribution from the disk at this stage with respect to the rising hard state. While Epochs 30 and 31 are characterised by a cold ($kT_{\rm disk}\sim0.3$ keV) disk, whose inner radius is unfortunately very loosely constrained, in the last two Epochs, the contribution from both the disk and the reflection components were not found to be statistically required by the fit. 

We suggest that, as the source evolves towards quiescence, the disk recedes and cools down significantly. Alternatively, the reduced disk contribution observed in Phase 4 may also be related to the progressive depletion of the outer accretion disk, leading to a weaker and negligible disk emission. Both scenarios are consistent with the observed X-ray spectral properties of the source in quiescence, where the source spectrum consists only of thermal emission from the NS surface and often a non-negligible contribution from a non-thermal, power-law component of debated origin \citep[e.g.][]{Chakrabarty2014,Cotizelati2014, Waterhouse2016,Ootes2018, Marino2018}.

\section{Conclusions}\label{sec:concl}
In this manuscript, we have presented an analysis of an extensive X-ray (including \ep, \nicer, \nustar) and optical/UV (\lco, \swift) campaign targeting the archetypal NS LMXB \src during its 2024 bright outburst.  Our results allow us to estimate the time-scales for the rise-phase of the outburst, in particular the start time for the outburst at different electromagnetic bands, the optical-to-X-ray delay for the outburst early rise and the time required to transition from the hard to the soft state. Additionally, with spectral-timing analysis, we were able to reconstruct in detail the evolution of the main physical parameters of the accretion flow throughout the event. Our main conclusions are the following:
\begin{itemize}
    \item Using both Gaussian Process modeling and a phenomenological analysis, we estimate that the 2024 outburst began around MJD 60543 in the optical band, while the soft X-ray onset occurred after MJD 60556, implying an optical-to-X-ray delay of $\lesssim$13 days. 
    \item During its 50-day outburst, \src displayed all the three canonical XRB spectral states: the hard, intermediate, and soft states. Throughout the outburst, the spectra are well described by three main components: a disk blackbody, an additional blackbody likely coming from the NS surface and Compton-scattered by a hot inner flow or corona and a reflection component. 
    \item Our spectral analysis provided estimates on the main spectral parameters during both the hard and soft states, tracing their evolution throughout the outburst. In particular: the electron temperature of the hot corona goes from $\sim$23 keV to $\sim$3 keV; the temperatures of the blackbody and disk components go from $\sim$0.6 to $\sim$1.6 keV and from $\sim$0.3 to $\sim$0.9 keV, respectively; the black body size shrinks from an initial value of $\sim$11 km to a final range between $\sim$3-8 km in the soft state; the inner radius of the disk goes from an initial value of about $\sim$60-100 km to $\sim$ 10 km (although these values are likely underestimated).
    \item Along with the spectral parameters, the fractional rms also shows a significant change between the two states, going from an initial $\sim$30-40\% to $\lesssim$10\% in the soft state. 
    \item Two different \texttt{relxill} code \citep{Garcia2013} versions were used to model the reflection component, \texttt{relxillCp} in the hard and \texttt{relxillNS} in the soft state, consistently with the expected drastic evolution of the Comptonization component spectral shape in the two states.
    \item Our dense monitoring covered with almost unprecedented detail the hard-to-soft transition, occurred between MJD 60580 and 60582, indicating it happened in two days, with the last dramatic evolution within just 12 hours (phase ``2b''). 
    \item We observed a peculiar increasing trend in the disk normalization during the day leading up to the hard-to-soft transition. While such an evolution could suggest a fluctuation of the inner disk edge around the transition, we suggest an alternative scenario where as the mass-accretion rate increases, a geometrically thicker inner disk emerges. In this scenario, the increase in normalization reflects the increase in the visible emitting area of a disk ``puffed up'' because of an increasing contribution by advection. 
\end{itemize}
LMXBs have been extensively studied for decades, yet we still poorly understand the physical mechanisms driving the onset and the evolution of their outbursts, and the role, if any, that the nature of the accreting star plays in this context. In this work, we have for the first time monitored the early phase of an outburst by \src, capturing the source at X-ray luminosity of less than 10$^{35}$ \ergs. In the EP era, LMXB outbursts will be identified at analogously low luminosities, providing not only dense X-ray monitoring of the early outbursts' rise but also the chance to trigger follow-up observations at other wavelengths, such as radio or optical, in this regime. Future multi-band campaigns driven by EP have the chance to open a new window on the behaviour of LMXBs at low accretion rates, thereby increasing our understanding of the accretion history of these sources.

\begin{acknowledgements}
We thank the anonymous referee for their constructive feedback and useful suggestions, which allowed us to significantly improve the scientific solidity of this work. We additionally thank V. Grinberg for useful discussion. This work makes use of observations from the Las Cumbres Observatory global telescope network. AM, EP, FCZ, MI and NR are supported by a ERC Consolidator Grant “MAGNESIA” under grant agreement No. 817661 (PI: Rea) and the National Spanish grant PID2023-153099NA-I00 (PI: Coti Zelati).
FCZ is supported by a Ram\'on y Cajal fellowship (grant agreement RYC2021-030888-I). Y.C. acknowledges support from the grant RYC2021-032718-I, financed by MCIN/AEI/10.13039/501100011033 and the European Union NextGenerationEU/PRTR and funds from the Spanish MINECO (PID2023-148661NB-I00)/E.U. FEDER (PI Jos\'e). NR and MI are also supported by a ERC Proof of Concept Grant “DeepSpacePulse” under grant agreement No. 101189496 (PI: Rea). MCB acknowledges support from the INAF-Astrofit fellowship and from the INAF Fundamental Research Grant XBOOM. This work was also partially supported by the program Unidad de Excelencia Maria de Maeztu CEX2020-001058-M. DMR is supported by Tamkeen under the NYU Abu Dhabi Research Institute grant CASS. RS acknowledges the INAF grant No. 1.05.23.04.04. GI is supported by a Juan de La Cierva fellowship (JDC2024-053550-I). YFH is supported by the National Key R\&D Program of China (2021YFA0718500) and by the Xinjiang Tianchi Program.AP acknowledges support from grant  PID2024-155316NB-I00, PID2021-124581OB-I0, and 2021SGR00426. GZ acknowledges support from the China Manned Space Program with grant No. CMS-CSST-2025-A13. 

\end{acknowledgements}

\bibliographystyle{aa}
\bibliography{biblio}

\appendix
\section{Tables and additional figures}
\begin{figure}
\centering
\includegraphics[scale=0.32]{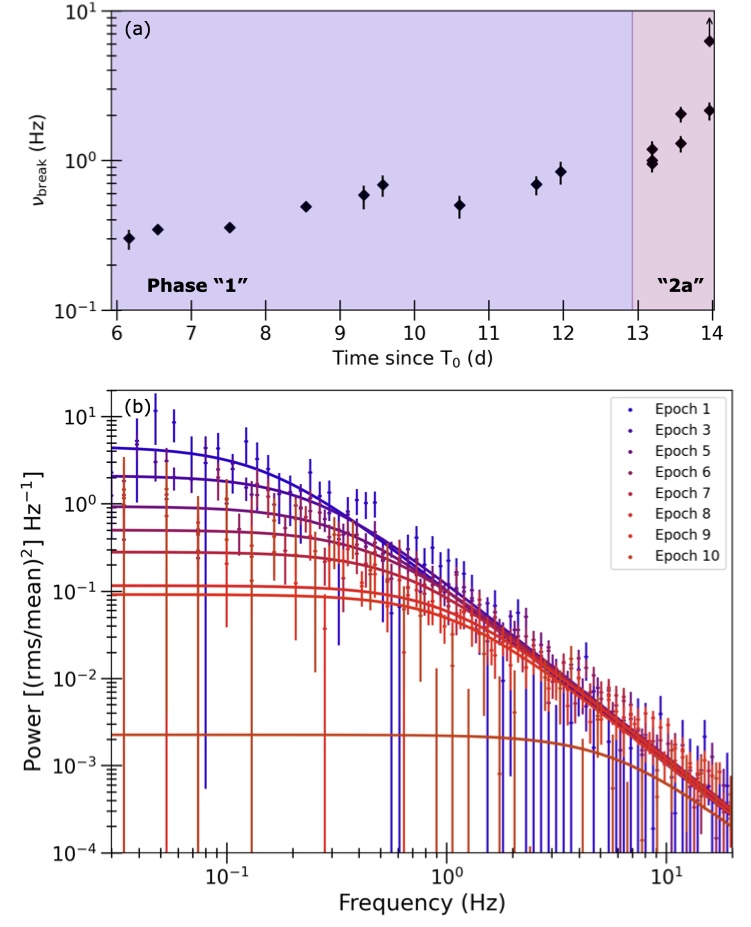}
\caption{Panel (a): evolution of the break frequency $\nu_{\rm break}$ throughout the hard to intermediate states in \src, obtained from the analysis of single \nicer PDS spectra; panel (b) selected PDS and best-fit single Lorentzian models, showing the increase in $\nu_{\rm break}$ over time.} 
\label{fig:nu_break}
\end{figure}

In this Section, we report details and results of this work in several Tables and Figures. In particular, Table \ref{tab:spec-log} lists all the (X-ray) observations and illustrates the Epochs classification we adopted, Table \ref{tab:multiband} reports the results of the phenomenological analysis explained in Section \ref{ss:lcurves}, Table \ref{tab:fit_broadband} shows the results of the fit to the broadband \nicer+\nustar spectra in Epochs 3 and 28, while Table \ref{tab:individual-fits} reports the spectral analysis on each individual Epoch (more details in Section \ref{ss:spectral}). Finally, the evolution of $\nu_{\rm break}$ through the hard and hard-intermediate states (Phases ``1'' and ``2a''), obtained through the analysis described in Section \ref{ss:timing}, is displayed in Fig. \ref{fig:nu_break}.

\begin{table*}[]
            \centering
            \caption{Table of the X-ray observations used in this work.}
            \begin{tabular}{l l l l l l }

\hline 
\hline 
 Observation ID & Telescope/Instrument &  Observation start (MJD) & Observation start (UTC) & Exposure (ks) & Epoch \T \B \\  
\hline 
7050340101 & NICER & 60573.664 & 2024-09-20 15:55:57.184 & 1.7 & \multirow{3}{*}{Epoch 1 }  \T  \\ 
7050340102 & NICER & 60574.049 & 2024-09-21 01:11:14.184 & 3.7 &  \\ 
06800000116 & EP/FXT  & 60574.067 & 2024-09-21 01:36:54.067 & 4.2 &  \B  \\ 
\hline 
7050340103 & NICER & 60575.015 & 2024-09-22 00:20:58.184 & 5.5 & \multirow{2}{*}{Epoch 2 }  \T  \\ 
06800000119 & EP/FXT  & 60575.337 & 2024-09-22 08:04:53.088 & 5.5 &  \B  \\ 
\hline 
91001338002 & NuSTAR & 60576.591 & 2024-09-23 14:10:48.684 & 18.3 & \multirow{5}{*}{Epoch 3 }  \T  \\ 
7050340104 & NICER & 60576.045 & 2024-09-23 01:04:16.184 & 4.1 &  \\ 
7675010101 & NICER & 60576.818 & 2024-09-23 19:37:59.184 & 0.6 &  \\ 
7050340105 & NICER & 60577.012 & 2024-09-24 00:17:03.184 & 0.3 &  \\ 
7675010102 & NICER & 60577.077 & 2024-09-24 01:51:11.184 & 0.9 &  \B  \\ 
\hline 
06800000122 & EP/FXT  & 60577.755 & 2024-09-24 18:07:28.974 & 3.5 & {Epoch 4 }  \T  \B  \\ 
\hline 
7675010103 & NICER & 60578.560 & 2024-09-25 13:27:02.184 & 0.2 &  \multirow{5}{*}{Epoch 5 } \T \\ 
7050340107 & NICER & 60579.141 & 2024-09-26 03:23:19.184 & 1.2 &    \\ 
7675010104 & NICER & 60579.464 & 2024-09-26 11:07:55.184 & 1.0 &  \\ 
06800000124 & EP/FXT  & 60579.479 & 2024-09-26 11:29:09.069 & 4.9 &    \\ 
7675010105 & NICER & 60580.045 & 2024-09-27 01:04:11.184 & 0.3 & \B \\ 
\hline 
7050340108$\dagger$ & NICER & 60580.690 & 2024-09-27 16:33:25.702 & 1.0 & \multirow{3}{*}{Epoch 6 }  \T  \\ 
7050340108$\dagger$  & NICER & 60580.755 & 2024-09-27 18:06:43.184 & 1.0 &  \\ 
7050340108$\dagger$  & NICER & 60580.883 & 2024-09-27 21:12:08.184 & 0.9 &  \B  \\ 
\hline 
7675010106$\dagger$  & NICER & 60581.077 & 2024-09-28 01:51:10.337 & 0.6 & {Epoch 7 }  \T  \B  \\ 
\hline 
7675010106$\dagger$  & NICER & 60581.335 & 2024-09-28 08:02:50.184 & 1.0 & {Epoch 8 }  \T  \B  \\ 
\hline 
7050340109$\dagger$  & NICER & 60581.465 & 2024-09-28 11:08:58.389 & 0.8 & {Epoch 9 }  \T  \B  \\ 
\hline 
7050340109$\dagger$  & NICER & 60581.668 & 2024-09-28 16:02:08.184 & 0.2 & {Epoch 10 }  \T  \B  \\ 
\hline 
06800000125$\dagger$  & EP/FXT  & 60581.692 & 2024-09-28 16:37:05.989 & 2.2 & {Epoch 11 }  \T  \B  \\ 
\hline 
06800000125$\dagger$  & EP/FXT  & 60581.761 & 2024-09-28 18:15:44.977 & 2.0 & {Epoch 12 }  \T  \B  \\ 
\hline 
7050340109$\dagger$  & NICER & 60581.916 & 2024-09-28 21:59:17.184 & 1.9 & {Epoch 13 }  \T  \B  \\ 
\hline 
7050340110$\dagger$  & NICER & 60582.240 & 2024-09-29 05:45:49.069 & 0.8 & {Epoch 14 }  \T  \B  \\ 
\hline 
7675010107$\dagger$  & NICER & 60582.368 & 2024-09-29 08:49:21.367 & 1.1 & {Epoch 15 }  \T  \B  \\ 
\hline 
7675010107$\dagger$  & NICER & 60582.626 & 2024-09-29 15:01:16.184 & 1.0 & {Epoch 16 }  \T  \B  \\ 
\hline 
7050340110$\dagger$  & NICER & 60582.755 & 2024-09-29 18:06:54.184 & 1.0 & {Epoch 17 }  \T  \B  \\ 
\hline 
7050340110$\dagger$  & NICER & 60582.834 & 2024-09-29 20:00:37.184 & 1.1 & {Epoch 18 }  \T  \B  \\ 
\hline 
7050340111$\dagger$  & NICER & 60583.077 & 2024-09-30 01:51:32.191 & 0.4 & {Epoch 19 }  \T  \B  \\ 
\hline 
7050340111$\dagger$  & NICER & 60583.207 & 2024-09-30 04:57:53.184 & 2.4 & \multirow{2}{*}{Epoch 20 }  \T  \\ 
7675010108 & NICER & 60583.400 & 2024-09-30 09:36:15.184 & 2.2 &  \B  \\ 
\hline 
7675010109 & NICER & 60584.756 & 2024-10-01 18:08:04.184 & 2.1 & {Epoch 21 }  \T  \B  \\ 
\hline 
7050340113 & NICER & 60585.014 & 2024-10-02 00:20:10.184 & 1.7 & {Epoch 22 }  \T  \B  \\ 
\hline 
7675010110 & NICER & 60585.853 & 2024-10-02 20:28:14.184 & 2.0 & {Epoch 23 }  \T  \B  \\ 
\hline 
7675010111 & NICER & 60586.305 & 2024-10-03 07:18:38.184 & 2.2 & \multirow{2}{*}{Epoch 24 }  \T  \\ 
7050340114 & NICER & 60586.692 & 2024-10-03 16:36:43.184 & 1.2 &  \B  \\ 
\hline 
7675010112 & NICER & 60587.209 & 2024-10-04 05:01:14.184 & 1.3 & {Epoch 25 }  \T  \B  \\ 
\hline 
7675010113 & NICER & 60588.437 & 2024-10-05 10:28:45.184 & 2.2 & {Epoch 26 }  \T  \B  \\ 
\hline 
7675010118 & NICER & 60593.413 & 2024-10-10 09:54:11.184 & 0.4 & {Epoch 27 }  \T  \B  \\ 
\hline 
7675010119 & NICER & 60594.123 & 2024-10-11 02:57:47.184 & 0.4 &\multirow{2}{*}{Epoch 28 }  \T  \\ 
91001345002 & NuSTAR & 60602.114 & 2024-10-19 02:43:53.684 & 20.1 &  \B  \\ 
\hline 
7675010133 & NICER & 60616.708 & 2024-11-02 16:59:05.184 & 1.0 & \multirow{2}{*}{Epoch 29 }  \T  \\ 
7675010134 & NICER & 60617.096 & 2024-11-03 02:18:23.184 & 1.9 &   \B  \\ 
\hline 
7050340124 & NICER & 60618.513 & 2024-11-04 12:19:21.184 & 1.9 & {Epoch 30 }  \T  \B  \\ 
\hline 
7675010135 & NICER & 60621.090 & 2024-11-07 02:10:00.184 & 2.2 & {Epoch 31 }  \T  \B  \\ 
\hline 
7675010136 & NICER & 60622.187 & 2024-11-08 04:28:51.184 & 1.5 & {Epoch 32 }  \T  \B  \\ 
\hline 
7675010137 & NICER & 60623.154 & 2024-11-09 03:42:11.184 & 2.1 & {Epoch 33 }  \T  \B  \\ 
\hline 
\hline 
\end{tabular}\label{tab:spec-log}
\tablefoot{All the EP/FXT observations used in this work were performed in TM mode. $\dagger$: The observation has been split into segments because of intra-observational variability. }
\end{table*}

\begin{table*}
\centering
\caption{Results of the multi-band light curves analysis}
\begin{tabular}{ l l l l l l l}
\hline 
\hline
  \multicolumn{6}{c}{\bf Multi-band phenomenological analysis} \T \B \\
\hline
\hline
Instrument/Filter & Energy range & & & & \T \B \\
  \hline
  &  & \multicolumn{4}{c}{X-ray flux, $f_\mathrm{X}$ (Broken exponential)} \T  \\
  & (keV) & $A_X$ (cts/s) & $\alpha_1$ & $\alpha_2$ & $T_{\mathrm{start, X}}$ (d) & $T_{\mathrm{knee}}$ (d) \B \\
  \hline
EP/WXT & 0.5-4 & 0.0150 $\pm$ 0.0012 & -0.50$\pm$0.02 & -0.121$\pm$0.003 & $>-11.0$ & 4.9$\pm$0.1 \T \\
\maxi & 3-20 & 0.004$\pm$0.003 & -0.70 $\pm$ 0.14 & -0.050$\pm$0.017 & $>-6.1$ & 5.7$\pm$0.3 \B \\ 
\hline
&   & \multicolumn{4}{c}{Optical magnitude, $\mathrm{mag}_\mathrm{opt}$ (Piecewise linear)} \T  \\
  & (\AA{}) & $\mathrm{mag}_\mathrm{Q, apparent}$ & $\mathrm{mag}_\mathrm{Q, real}$ & $a_\mathrm{opt}$ & $T_{\mathrm{start, opt}}$ (d) \B \\
  \hline
LCO/z' & 7660-9740 & 17.920 $\pm$ 0.009 & - & -0.0970 $\pm$ 0.0021 & -  \T \\
LCO/i' & 6255-8835 & 18.200$\pm$0.008 & (20.3) & -0.1068 $\pm$ 0.0015 & $>-24.1$ &    \\ 
LCO/R  & 4827-7987 &  18.42$\pm$0.02 & -& -0.1140$\pm$0.0017 & - &   \\ 
LCO/r' & 4825-7605 &  18.770$\pm$0.009 & - & -0.140$\pm$0.004 & - & \\ 
LCO/V  & 4608-6288 & 19.26$\pm$0.04 & (21.6) & -0.130$\pm$0.003 &  $>-23.7$ \\ 
LCO/g' & 3270-6270 &  19.73$\pm$0.02 & -& -0.150$\pm$0.003 & -  \B \\ 
\hline
& & \multicolumn{4}{c}{Ultra-violet magnitude, $\mathrm{mag}_\mathrm{UV}$ (Simple linear)} \T  \\
  & (\AA{})$^\ddagger$ & & & $a_\mathrm{UV}$  \B \\
  \hline
  UVOT/B & 3417-5367 & & & -0.11$\pm$0.02 \T \\ 
  UVOT/V & 4699-6237 & & & -0.11$\pm$0.02  \B \\
  
\hline
\hline
\end{tabular}
\tablefoot{All times are reported with respect to $T_0=60567.5$ MJD. See Section \ref{ss:lcurves} for more details on the employed modeling.  $^\ddagger$: nominal bands from \cite{Poole2008}}
\label{tab:multiband}
\end{table*}

\begin{table*}
\caption{Results of the broadband spectral analysis.}
\centering
\begin{tabular}{ l l l p{4 cm} l l}
\hline 
\hline
  \multicolumn{5}{c}{\bf Broadband spectral analysis} \T \B \\
\hline
\hline
  Component & \multicolumn{2}{c}{\bf Parameters} & Description & Epoch 3 & Epoch 27 \T \B \\
  \hline
\texttt{constant} & $c_{\rm cal}$ & & Intercalibration constant & \multicolumn{2}{c}{(1.0)} \\ \T \B \\
\hline
\texttt{TBabs} & $N_H$ & 10$^{22}$ cm$^{-2}$ & Equivalent hydrogen column density & \multicolumn{2}{c}{(0.50)} \\ \T \B \\
\hline
\multirow{3}{*}{\texttt{thComp}} & $\Gamma$ &  & Power-law index of the Comptonization spectrum & 1.85$\pm$0.01 & 3.3$^{+0.3}_{-0.2}$\T \\ 
& $kT_e$ & keV & Electron temperature of the corona & 23$^{+2}_{-1}$ & 3.4$^{+0.3}_{-0.2}$ \\
& $f_{\rm cov}$ &  & Covering fraction & \multicolumn{2}{c}{(1.0)} \\ \B\\
\hline
\multirow{2}{*}{\texttt{bbodyrad}} & $kT_{\rm bb}$ & keV & Blackbody temperature & 0.72$^{+0.03}_{-0.02}$ & 1.50$^{+0.06}_{-0.03}$\T \\
& $R_{\rm bb}$ & km & Blackbody radius$^\dagger$ & 11.1$^{+0.7}_{-0.8}$ & 4.4$^{+0.2}_{-0.4}$ \B\\
\hline
\multirow{2}{*}{\texttt{diskbb}} & $kT_{\rm disk}$ & keV & Inner disk temperature & 0.38$\pm$0.02 & 0.93$^{+0.05}_{-0.04}$  \T \\
& $R_{\rm disk}$ & (km) & Disk inner radius$^\dagger$ & 14.4$^{+0.8}_{-0.6}$ & 8.9$^{+1.0}_{-0.8}$ \B \\ 
\hline
\texttt{expabs} & $E_{\rm cut-off, low}$ & keV & Low energy cut-off & =3$\times kT_{\rm bb}$ & - \T \B \\ 
\hline
\multirow{12}{*}{\texttt{relxillCp}} & $i$ & $^\circ$ & System inclination & 30$\pm$2 & - \T \\ 
& $a*$ &  & Spin parameter & (0) & - \\ 
& $R_{\rm in}$ & $R_{\rm G}$ & Inner disk radius & $<$8.0 & - \\ 
& $R_{\rm out}$ & $R_{\rm G}$ & Outer disk radius & (1000)   - \\
& $\epsilon$ &  & Disk emissivity  & (3.0) & - \\
& $z$ &  & Redshift to the source & (0)  & - \\
& $\Gamma_{\rm relxill}$ &  & Power-law index of the incident spectrum & =$\Gamma$ & - \\ 
& $\log{\xi}$ &  & Disk ionization  & 3.00$^{+0.10}_{-0.12}$  & - \\
& $\log{N}$ & cm$^{-3}$ & Disk density & (19.0) & - \\
& A$_{\rm Fe}$ &  & Fe abundance of reflecting material & (1.0) & - \\
& $kT_{\rm e, refl}$ & keV & Electron temperature of the corona & =$kT_{\rm e}$ & - \\
& $f_{\rm refl}$ &  & Reflection fraction & (-1.0) & - \\ 
& $K_{\rm refl}$ & ($\times$10$^{-3}$)  & Reflection normalization & 2.30$\pm$0.17 & - \B \\
\hline
\multirow{12}{*}{\texttt{relxillNS}} & $i$ & $^\circ$ & System inclination & - & 31$\pm$2 \T \\ 
& $a*$ &  & Spin parameter & - & (0) \\ 
& $R_{\rm in}$ & $R_{\rm G}$ & Inner disk radius & - & $13^{+5}_{-4}$ \\ 
& $R_{\rm out}$ & $R_{\rm G}$ & Outer disk radius & - & (1000) \\
& $\epsilon$ &  & Disk emissivity  & - & (3.0) \\
& $z$ &  & Redshift to the source & - & (0) \\
& $kT_{\rm bb, relx}$ & keV  & Blackbody temperature of the incident spectrum & - & =$kT_{\rm bb}$ \\ 
& $\log{\xi}$ &  & Disk ionization  & - & 3.01$^{+0.12}_{-0.06}$\\
& $\log{N}$ & cm$^{-3}$ & Disk density & - & (19.0) \\
& A$_{\rm Fe}$ &  & Fe abundance of reflecting material & - & (1.0) \\
& $kT_{\rm e, refl}$ & keV & Electron temperature of the corona & - & =$kT_{\rm e}$ \\
& $f_{\rm refl}$ &  & Reflection fraction & - & (-1.0) \\ 
& $K_{\rm refl}$ & ($\times$10$^{-3}$)  & Reflection normalization & - & 2.2$^{+0.3}_{-0.2}$ \B \\
\hline
\texttt{cflux} & $F_{1-10 \ {\rm keV}}$ & \tiny{($\times$10$^{-9}$) \flux} & X-ray unabsorbed flux & 2.290$\pm$0.004 & 11.80$\pm$0.01  \T \B \\
\hline
& $\chi^2$ & (dof) & &  892(960) & 298(309) \T \B \\  
\hline
\hline
\end{tabular}
\tablefoot{Quoted errors reflect 90\% confidence level. The parameters that were kept frozen during the fits are reported between round parentheses. $R_{\rm G}$ represents the gravitational radius. The reported flux values correspond to the 1--10~keV energy range. $\dagger$: estimated from the parameter normalization and likely underestimated (see text for more details).}
\label{tab:fit_broadband}
\end{table*}

\begin{sidewaystable*}[]
            \centering
            \caption{Results of the spectral analysis for each individual Epoch.}
            \begin{tabular}{l l l l l l l l l l l }
\hline 
\hline 
Epoch & Time & N$_H$ & $\Gamma$ & kT$_{\rm bb}$ & R$_{\rm bb}$ & kT$_{\rm disk}$ & R$_{\rm disk}$ & K$_{\rm refl}$ & L$_{110}$ & $\chi^2$ \T \B \\ 
 & (d) & ($\times$10$^{22}$) & & (keV) & (km) & (keV) & (km) &  & ($\cdot$10$^{36}$ erg/s) & (d.o.f.) \T \B \\ 
\hline 
1 & 60573.93 & (0.500) & 1.85$\pm$0.03& 0.56$\pm$0.02& 13.0$^{+1.0}_{-0.9}$& 0.20$\pm$0.02& 50$^{+14}_{-8}$& 0.95$\pm$0.24 & 3.420$\pm$0.013& 362(370)\T \B \\  
2 & 60575.18 & (0.500) & 1.81$\pm$0.03& 0.62$\pm$0.03& 12.5$^{+1.1}_{-1.0}$& 0.30$\pm$0.02& 24$^{+5}_{-3}$& 1.9$^{+0.4}_{-0.3}$ & 4.890$\pm$0.013& 181(226)\T \B \\  
3 & 60576.69 & (0.500) & 1.850$^{+0.011}_{-0.008}$& 0.71$\pm$0.03& 11.0$^{+0.7}_{-0.8}$& 0.40$\pm$0.02& 14.4$^{+0.8}_{-0.6}$ & 2.3$\pm$0.2& 6.805$\pm$0.013& 887(961)\T \B \\  
4 & 60577.76 & (0.500) & 1.80$^{+0.12}_{-0.10}$& 0.60$\pm$0.09& 15.0$^{+6.0}_{-4.0}$& 0.25$\pm$0.03& 39$^{+12}_{-6}$& 4.0$^{+1.9}_{-1.6}$& 7.40$\pm$0.03& 106(77)\T \B \\  
5 & 60579.34 & (0.500) & 1.80$\pm$0.03& 0.72$\pm$0.03& 12.0$^{+1.1}_{-0.9}$& 0.35$\pm$0.01& 19.5$\pm$0.9 & 4.4$\pm$0.5& 8.95$\pm$0.03& 568(590)\T \B \\  
6 & 60580.78 & (0.500) & 1.80$\pm$0.03& 0.75$\pm$0.04& 12.5$^{+1.3}_{-1.1}$& 0.40$\pm$0.02& 22.5$^{+1.3}_{-1.1}$& 2.0$^{+1.1}_{-1.2}$& 12.00$\pm$0.04& 354(400)\T \B \\  
7 & 60581.08 & (0.500) & 1.75$\pm$0.05& 0.70$\pm$0.05& 16$^{+3}_{-2}$& 0.34$\pm$0.02& 29$^{+3}_{-2}$& 8.5$\pm$1.3& 14.00$\pm$0.05& 131(129)\T \B \\  
8 & 60581.34 & (0.500) & 1.75$\pm$0.04& 0.70$\pm$0.04& 17.5$^{+2.0}_{-1.7}$& 0.35$\pm$0.01& 32.5$^{+1.3}_{-1.2}$&    -     & 15.50$\pm$0.05& 115(136)\T \B \\  
9 & 60581.46 & (0.500) & 1.75$\pm$0.04& 0.70$\pm$0.05& 17.5$^{+2.5}_{-2.0}$& 0.35$\pm$0.01& 30.1$^{+1.3}_{-1.1}$ &    -     & 16.10$\pm$0.05& 108(134)\T \B \\  
10 & 60581.67 & (0.500) & 1.75$^{+0.30}_{-0.20}$& 1.30$^{+0.40}_{-0.14}$& 5.0$\pm$1.5& 0.60$^{+0.07}_{-0.02}$& 15.5$^{+0.8}_{-1.5}$ & 1.50$\pm$0.46 & 19.65$\pm$0.09& 104(119)\T \B \\  
11 & 60581.69 & (0.500) & (1.600) & 1.60$^{+0.13}_{-0.30}$& 3.0$^{+3.0}_{-0.7}$& 0.690$^{+0.020}_{-0.050}$ & 12.5$^{+1.4}_{-0.4}$& 2.4$^{+0.8}_{-1.1}$ & 21.20$\pm$0.05& 118(80)\T \B \\  
12 & 60581.76 & (0.500) & (1.600) & 1.3$\pm$0.3& 5.0$^{+7.0}_{-1.4}$& 0.75$\pm$0.06& 11.5$^{+1.4}_{-0.8}$& 2.2$^{+0.6}_{-0.9}$ & 26.70$\pm$0.06& 91(83)\T \B \\  
13 & 60581.92 & (0.500) & 2.5$^{+0.4}_{-0.2}$ & 1.40$^{+0.16}_{-0.13}$ & 5.5$^{+1.5}_{-1.1}$ & 0.85$^{+0.05}_{-0.04}$ & 11.0$^{+0.6}_{-0.5}$ & 1.5$\pm$0.3 & 32.60$\pm$0.08 & 54(146)\T \B \\  
14 & 60582.24 & (0.500) & (3.100) & 1.25$\pm$0.07 & 7.2$^{+1.0}_{-0.9}$ & 0.80$\pm$0.04 & 11.5$\pm$0.6 & 1.2$\pm$0.3 & 35.80$\pm$0.10 & 83(140)\T \B \\  
15 & 60582.37 & (0.500) & 3.5$^{+2.0}_{-0.6}$ & 1.40$^{+0.18}_{-0.14}$ & 7.0$^{+1.9}_{-1.4}$ & 0.82$\pm$0.06 & 11.0$^{+0.9}_{-0.8}$ & 1.6$\pm$0.5 & 41.50$\pm$0.11 & 71(140)\T \B \\  
16 & 60582.63 & (0.500) & (3.100) & 1.30$\pm$0.07 & 8.0$\pm$1.0 & 0.80$^{+0.05}_{-0.04}$ & 11.6$\pm$0.7 & 1.7$\pm$0.5 & 44.90$\pm$0.12 & 57(141)\T \B \\  
17 & 60582.75 & (0.509) & 3.3$^{+1.3}_{-0.4}$ & 1.34$^{+0.17}_{-0.10}$ & 8.7$^{+1.7}_{-1.8}$ & 0.79$\pm$0.08 & 12.0$^{+1.0}_{-1.1}$ & 1.7$\pm$0.5 & 52.90$\pm$0.14 & 66(142)\T \B \\  
18 & 60582.83 & (0.500) & (3.100) & 1.38$\pm$0.10 & 7.6$^{+1.0}_{-1.2}$ & 0.87$\pm$0.07 & 10.3$\pm$0.7 & 2.0$^{+0.4}_{-0.5}$ & 49.60$\pm$0.12 & 68(147)\T \B \\  
19 & 60583.08 & (0.500) & (3.100) & 1.56$\pm$0.10 & 6.0$^{+1.2}_{-0.8}$ & 0.92$\pm$0.06 & 9.7$^{+0.8}_{-0.6}$ & 3.0$\pm$0.7 & 53.60$\pm$0.17 & 95(133)\T \B \\  
20 & 60583.30 & (0.500) & (3.100) & 1.34$\pm$0.03 & 8.3$\pm$0.5 & 0.82$\pm$0.03 & 11.3$\pm$0.4 & 1.7$\pm$0.3 & 49.20$\pm$0.11 & 124(302)\T \B \\  
21 & 60584.76 & (0.500) & (3.100) & 1.35$\pm$0.07 & 7.8$^{+0.8}_{-0.9}$ & 0.850$^{+0.050}_{-0.040}$ & 10.6$\pm$0.6 & 1.6$\pm$0.4& 47.50$\pm$0.10& 64(149)\T \B \\  
22 & 60585.01 & (0.500) & (3.100) & 1.35$\pm$0.08& 7.2$^{+0.9}_{-1.0}$& 0.82$^{+0.05}_{-0.04}$& 11.1$^{+0.6}_{-0.7}$ & 1.9$\pm$0.4& 42.90$\pm$0.10& 84(148)\T \B \\  
23 & 60585.85 & (0.500) & 3.0$^{+4.0}_{-0.7}$& 1.40$^{+0.19}_{-0.12}$& 5.8$^{+1.2}_{-1.1}$& 0.85$\pm$0.07& 10.5$^{+0.9}_{-0.7}$ & 3.4$^{+1.8}_{-1.6}$ & 42.50$\pm$0.10& 55(144)\T \B \\  
24 & 60586.50 & (0.500) & (3.100) & 1.30$\pm$0.04& 8.5$\pm$0.7& 0.80$\pm$0.03& 11.8$\pm$0.5 & 1.9$\pm$0.3& 47.15$\pm$0.12& 170(301)\T \B \\  
25 & 60587.21 & (0.500) & (3.100) & 1.41$\pm$0.09& 7.3$^{+0.9}_{-1.0}$& 0.85$^{+0.06}_{-0.04}$& 10.7$\pm$0.7 & 2.0$^{+0.4}_{-0.5}$ & 48.90$\pm$0.12& 86(147)\T \B \\  
26 & 60588.44 & (0.500) & (3.100) & 1.34$\pm$0.09& 7.8$^{+1.0}_{-1.1}$& 0.83$^{+0.06}_{-0.04}$& 10.9$\pm$0.7 & 2.0$\pm$0.4& 45.90$\pm$0.11& 95(149)\T \B \\  
27 & 60593.41 & (0.500) & 2.7$^{+2.9}_{-0.4}$& 1.5$^{+0.3}_{-0.2}$& 6$^{+3}_{-2}$& 0.92$\pm$0.09& 9.1$^{+1.3}_{-0.7}$ & 2.0$\pm$0.5& 44.10$\pm$0.13& 78(134)\T \B \\  
28 & 60599.45 & (0.500) & 3.3$^{+0.3}_{-0.2}$& 1.50$^{+0.06}_{-0.03}$& 4.4$^{+0.2}_{-0.4}$& 0.93$^{+0.05}_{-0.04}$& 8.9$^{+1.0}_{-0.8}$ & 2.2$^{+0.3}_{-0.2}$ & 35.31$\pm$0.04& 298(309)\T \B \\  
29 & 60616.90 & (0.500) & (3.100) & 1.19$\pm$0.03& 3.15$^{+0.12}_{-0.15}$& 0.530$\pm$0.010& 13.7$^{+0.4}_{-0.6}$ & - & 5.50$\pm$0.02& 308(259)\T \B \\  
30 & 60618.51 & (0.500) & 1.950$\pm$0.013& (0.220) & 60.8$^{+1.7}_{-1.6}$& 0.15$\pm$0.02& 74$^{+37}_{-14}$&    -     & 2.420$\pm$0.010& 87(127)\T \B \\  
31 & 60621.09 & (0.500) & 2.03$\pm$0.02& 0.34$\pm$0.02& 18.7$^{+1.0}_{-1.5}$& 0.15$\pm$0.03& 44$^{+44}_{-10}$&    -     & 0.950$\pm$0.005& 107(119)\T \B \\  
32 & 60622.19 & (0.500) & 2.21$\pm$0.03& 0.350$\pm$0.010 & 13.5$\pm$0.5&    -     &    -    &    -     & 0.511$\pm$0.004 & 93(89)\T \B \\  
33 & 60623.15 & (0.500) & 2.09$\pm$0.05& 0.33$\pm$0.01& 12.0$^{+0.7}_{-0.6}$&    -     &    -    &    -     & 0.323$\pm$0.003 & 82(67)\T \B \\  
\hline 
\hline 
\end{tabular}
\tablefoot{A distance of 4.5 kpc has been assumed to estimate the X-ray luminosity in the 1-10 keV range, $L_{110}$. The parameters that were kept frozen during the fits are reported between round parentheses.}
\label{tab:individual-fits}
\end{sidewaystable*}

\begin{figure}
\centering
\includegraphics[scale=0.4]{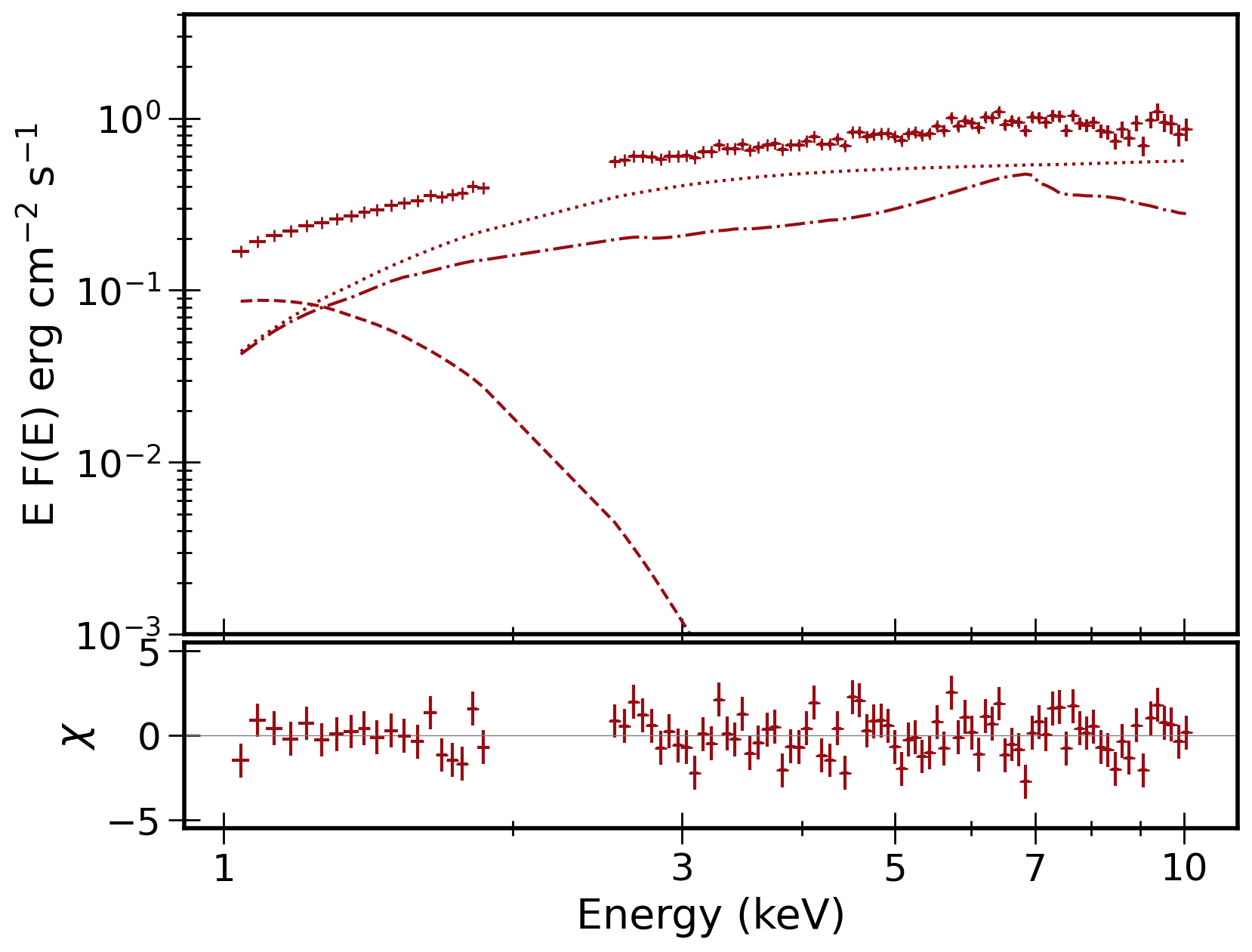}
\caption{ \ep/FXT spectrum for Epoch 4 fitted with Model H and residuals. Different line styles were adopted to distinguish between the different components: dash for \texttt{diskbb}, dot for \texttt{thcomp}$\times$\texttt{bbodyrad} and dash-dot for \textsc{relxillCp}. The 1.8-2.5 keV range has been ignored because of known FXT calibration issues for bright targets.} 
\label{fig:ep-spectrum}
\end{figure}

\begin{figure*}
\centering
\includegraphics[scale=0.3]{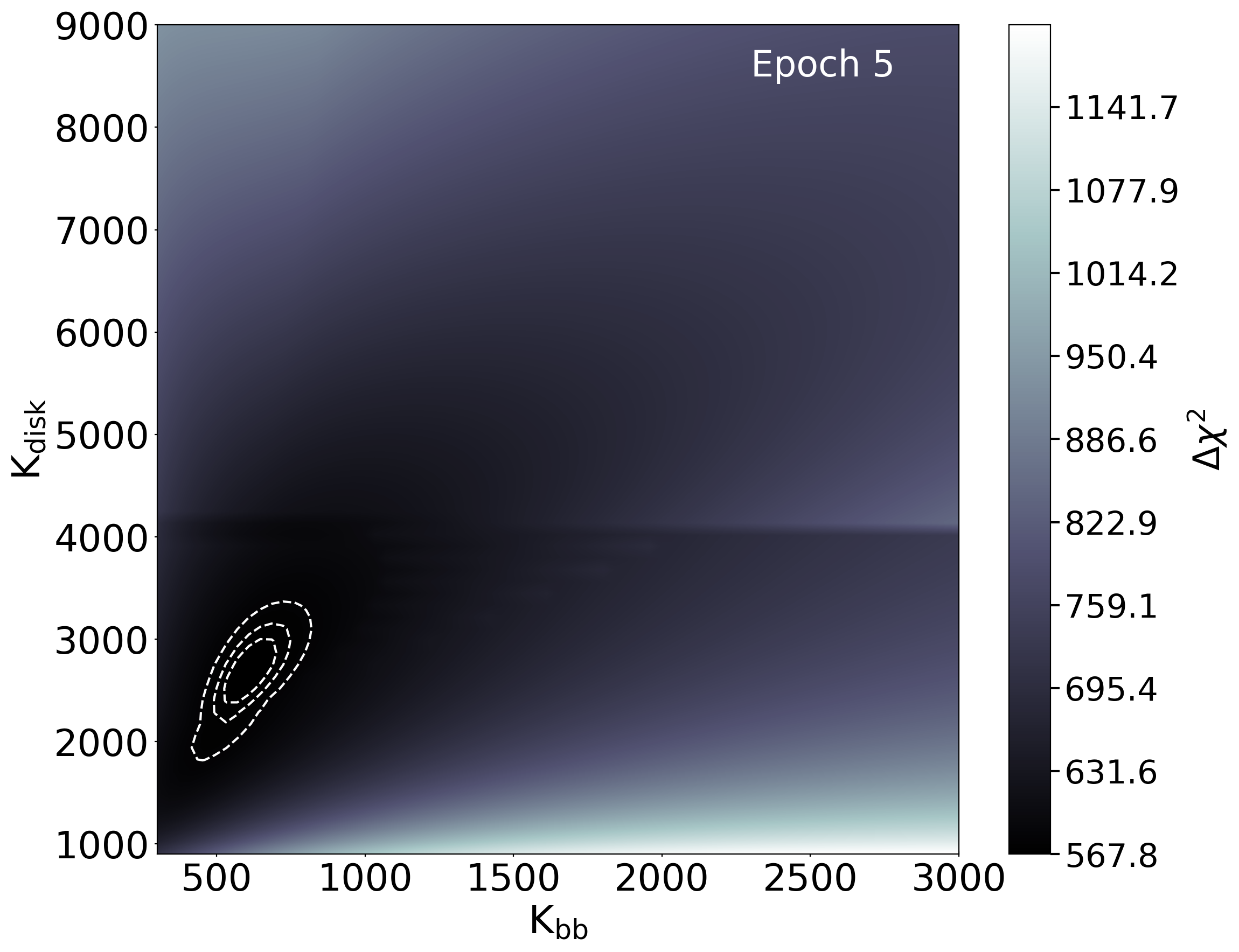}
\includegraphics[scale=0.3]{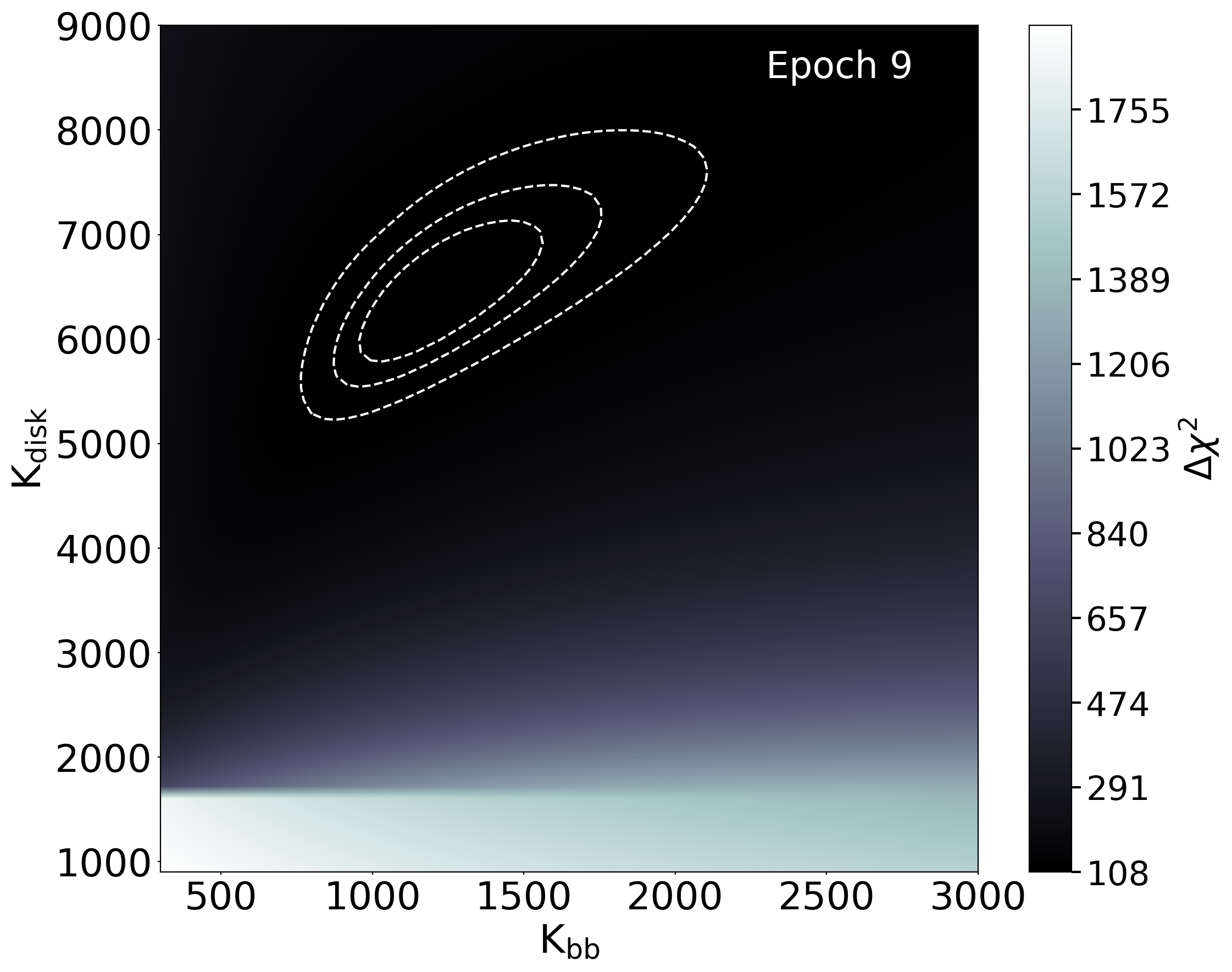}
\caption{ Contour plots of the disk and blackbody normalizations, $K_{\rm disk}$ and $K_{\rm bb}$, for Epoch 5 (left) and Epoch 9 (right). The dashed contours indicate the 99\%, 90\%, and 68\% confidence levels, from outermost to innermost. As apparent from the Figure, the lack of overlap between the confidence regions of each Epoch and the absence of secondary minima in the $\chi^2$ landscape indicate that the increase in both normalizations (and therefore radii) is statistically significant.} 
\label{fig:contour-plots}
\end{figure*}

\section{Early rise analysis with Gaussian Processes}\label{ss:GP}
We adopted the implementation provided in the \texttt{scikit-learn} library \citep{Pedregosa2011}, using a kernel composed of a constant term multiplied by a Matérn kernel with smoothness parameter $\nu = 1.5$, and added a white-noise component to account for measurement uncertainties. This configuration offers sufficient flexibility to capture both smooth trends and stochastic intrinsic variability.

We applied the Gaussian Process (GP) regression separately to the X-ray light curves from EP/WXT (0.5--4\,keV) and \maxi\ (3--20\,keV), as well as to the \lco\ optical data. To incorporate the two early EP/WXT upper limits without biasing the model, we introduced ``pseudo-measurements'' at half the reported upper-limit values, with uncertainties equal to half those values. This approach encodes the one-sided nature of the non-detections with a conservative 1$\sigma$ error while preventing the GP model from being over-constrained.

\begin{figure}
\centering
\includegraphics[scale=0.43]{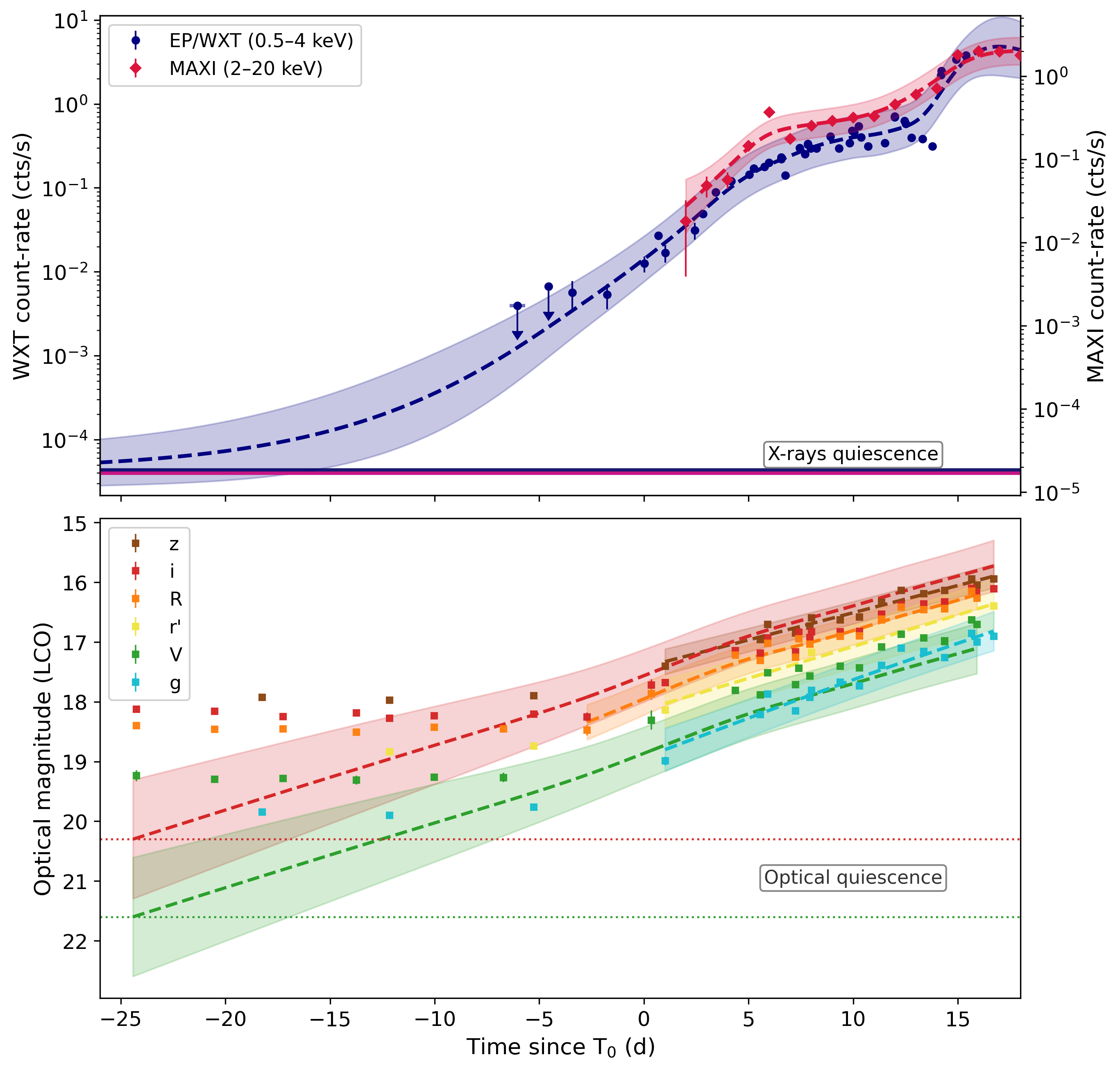}
\caption{
Multi-band light curves of Aql X-1 during the rise to outburst, spanning X-ray (top panel) and optical (bottom panel) observations, along with their respective GP fits and 1$\sigma$ confidence intervals. In the plot, the quiescent luminosity for EP/WXT, \maxi\ (top) and the quiescent magnitude for the i-band and V-band (bottom) are marked (see the caption for Fig. \ref{Fig:early-rise} for more details).} 
\label{Fig:early-rise-GP}
\end{figure}

Figure\,\ref{Fig:early-rise-GP} shows the GP models overlaid on the data, highlighting the outburst onset and early evolution across energy bands. We estimate the onset of the outburst as the latest time at which the outburst could have plausibly begun, identified as the last intersection between the upper bound of the 95\% confidence interval from the EP/WXT GP model and the estimated quiescent level, as determined via backward extrapolation. Based on this criterion, the enhanced activity likely began earlier than approximately MJD 60557, at $T_0-10 $ d, for the X-rays and earlier than MJD 60542, at $T_0-25 $ d, in the optical band. 

Also in this case, the above estimate relies on the smooth extrapolation provided by the GP model. Since the earliest EP/WXT measurements are upper limits, a sharp steepening or discontinuity in the X-ray flux just before the first detection cannot be excluded a priori. In that case, the actual outburst onset could have occurred later than inferred. Our estimate should therefore be interpreted under the assumption of a continuous rise and therefore as a lower limit for the true start of the outburst in the X-ray band.

\section{Spectral analysis with \texttt{diskpbb}}\label{ss:app-slim-disk}
\begin{figure}
\centering
\includegraphics[scale=0.32]{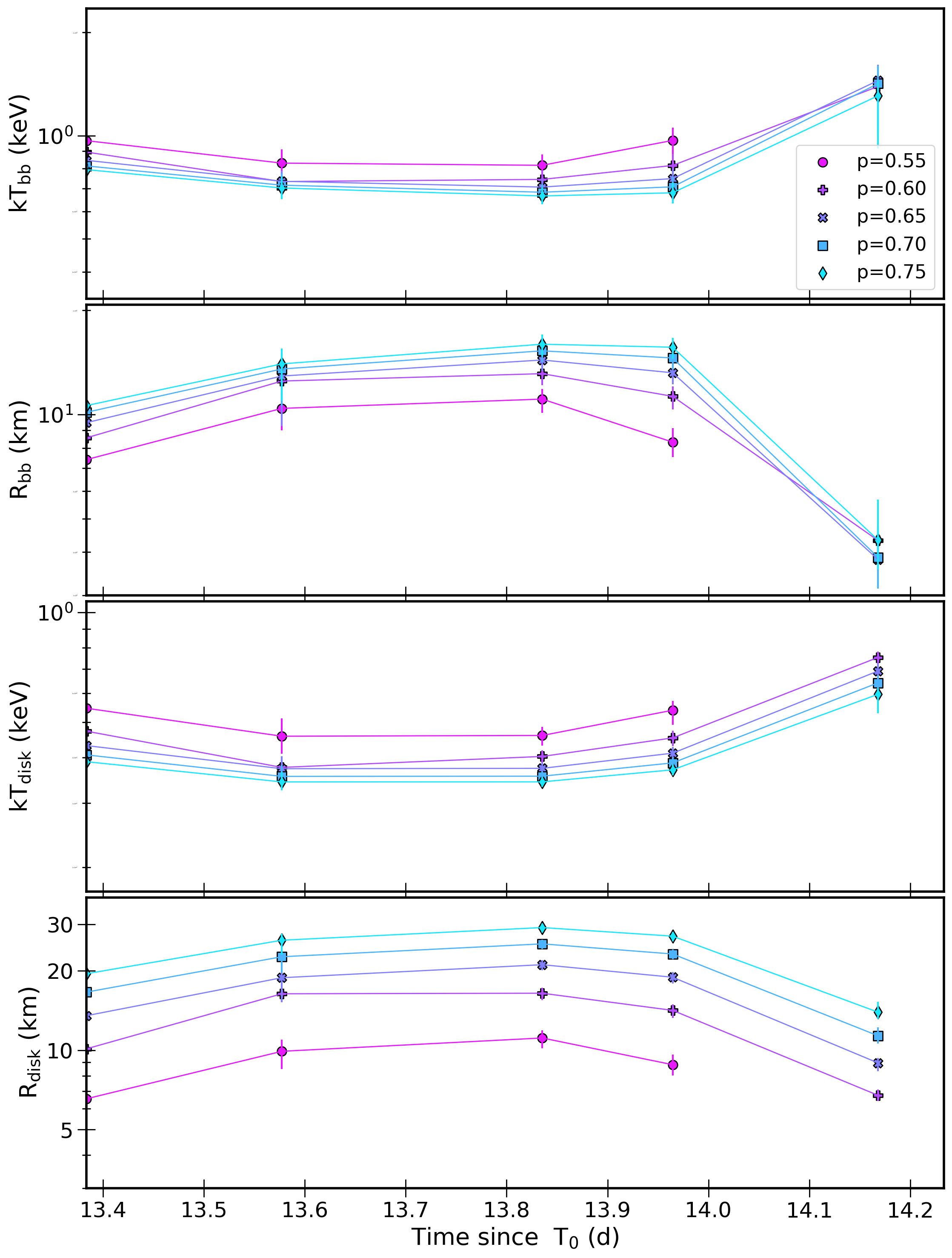}
\caption{Evolution of best-fit $kT_{\rm bb}$, $R_{\rm bb}$, $kT_{\rm disk}$ and $R_{\rm disk}$ for Epochs 6 to 10 using the \texttt{diskpbb} model and fixing the parameter $p$ to several values, going from 0.75 (classical \citealt{ShakuraSunyaev1973} disk, negligible energy losses through advection) to 0.5 (``slim'' disk, significant part of the energy lost through advection).}
\label{fig:slim-disk}
\end{figure}

In order to explore the possibility of an increase in scale height of the inner disk during the spectral state transition in \src, we have tested whether \texttt{diskpbb} might serve as a temporary replacement for \texttt{diskbb} for Epochs 6 to 9, where an odd increase in normalization was observed (see Discussion). For each of these Epochs, we have used the best-fit model identified in Section \ref{ss:spectral}, ``H'' (until Epoch 9) or ``S'' (Epoch 10), but replaced \texttt{diskbb} with \texttt{diskpbb} and re-run the fits. An initial attempt at running the fit with the radial temperature index $p$ as a free parameter was inconclusive, as the number of degeneracies in the fit prevented us from constraining this parameter. We therefore opted for a different approach, launching multiple fits with $p$ fixed to values between 0.55 and 0.75 in steps of 0.05, where $p=0.75$ corresponds to the standard thin-disk model of \citet{ShakuraSunyaev1973}. All fits are acceptable and comparable to the ones obtained with \texttt{diskbb}, as expected due to the degeneracies in the model used, with the exception of Epoch 10, for which using \texttt{diskpbb} with $p$ lower than 0.65 worsened the fit significantly. We show the evolution of $kT_{\rm bb}$, $R_{\rm bb}$ , $kT_{\rm disk}$ and $R_{\rm disk}$ in Fig. \ref{fig:slim-disk}. As evident from the plot, for all values of $p$ we retrieve the same trend for both parameters, with the radius (temperature) initially increasing (decreasing) to then decrease (increase) again. However, a clear correlation exists between the values of $p$ and both $R_{\rm bb}$ and $R_{\rm disk}$. If we therefore allow $p$ to slowly decrease from the initial thin-disk value (0.75) to a slightly lower value (0.55), we obtain an acceptable description of the spectra by keeping $R_{\rm disk}$ and $R_{\rm bb}$ fixed or even allowing them to decrease.

\end{document}